\documentstyle[preprint,aps,floats,epsfig]{revtex}

\tightenlines

\def\appendix{\par
 \setcounter{section}{0}
 \setcounter{subsection}{0}
 \def\thesection{Appendix \Alph{section}}
 \def\thesubsection{\Alph{section}.\arabic{subsection}}
 \def\theequation{\Alph{section}.\arabic{equation}}
 \setcounter{equation}{0}}
\newcommand{\be}{\begin{equation}}
\newcommand{\ee}{\end{equation}}
\newcommand{\bear}{\begin{eqnarray}}
\newcommand{\eear}{\end{eqnarray}}

\begin{document}
\draft
\title{Bilocal expansion of the Borel amplitude and \\
       the hadronic tau decay width\footnote{
To appear in Phys. Rev. D}}
\author{Gorazd Cveti\v c$^1$\footnote{cvetic@fis.utfsm.cl}
and Taekoon Lee$^2$\footnote{
tlee@muon.kaist.ac.kr}}
\address{$^1$Department of Physics, 
Universidad T\'ecnica Federico Santa Mar\'{\i}a,     
Valpara\'{\i}so, Chile\\
$^2$Department of Physics, KAIST, Taejon 305-701, Korea}
\maketitle
\begin{abstract}
The singular part of Borel transform of a QCD amplitude near the infrared
renormalon can be expanded in terms of higher order Wilson coefficients
of the operators associated with the renormalon. In this paper we observe
that this expansion gives nontrivial constraints on the Borel amplitude that
can be used to improve the accuracy of the ordinary perturbative expansion 
of the Borel amplitude. In particular, we consider the Borel transform 
of the Adler function and its expansion around the first infrared renormalon
due to the gluon condensate. Using the next--to--leading order Wilson 
coefficient of the gluon condensate operator, we obtain an exact constraint 
on the Borel amplitude at the first IR renormalon. We then extrapolate, 
using judiciously chosen conformal transformations and Pad\'e approximants, 
the ordinary perturbative expansion of the Borel amplitude in such a way 
that this constraint is satisfied. This procedure allows us to predict 
the ${\cal {O}}(\alpha_s^4)$ coefficient of the Adler function, 
which gives a result consistent with the 
estimate by Kataev and Starshenko using a completely different method. 
We then apply this improved Borel amplitude to the tau decay width,
and obtain the strong coupling constant $\alpha_{\text{s}}(M^2_{\text{z}})
=0.1193 \pm 0.0007_{\rm exp.} \pm 0.0010_{\rm EW+CKM} \pm 0.0009_{\rm meth.}
\pm 0.0003_{\rm evol.}$. 
We then compare this result with those of other resummation methods.

\end{abstract}
\section{Introduction}
\label{sec:introduction}

The ordinary perturbative expansion in quantum chromodynamics (QCD) gives
a divergent series, with rapidly increasing perturbative coefficients.
Having higher order corrections, thus, does not automatically mean better
accuracy. A further step should be taken to properly handle the divergent
series. For this purpose, the Borel resummation technique is often
invoked.

The Borel resummation  of the perturbation series in QCD, however, 
is not straightforward because of the nonperturbative effects that 
cause singularities on the Borel plane.
Generally, the Borel transform of QCD amplitude has singularities 
\cite{parisi,hooft},
the ultraviolet (UV) renormalons on the negative real axis, and
the infrared (IR) renormalons on the positive real axis. There are also 
singularities caused by instanton--anti-instanton pairs, but these are 
irrelevant to our discussion and  shall be ignored.

In Borel resummation the UV renormalons are not a serious problem,
since they can be 
transformed far away from the Borel integration contour using
  a proper conformal mapping, but the IR renormalons, which are
located on the integration contour, cause a real problem.
First of all, the IR renormalons cause ambiguities in taking a
proper contour at their positions. The IR renormalons can be
associated with certain operator condensates \cite{parisi} appearing
in operator product expansion,
and these ambiguities are 
known to arise from the ambiguities in defining the renormalized 
condensates in continuum limit \cite{david}. 
Because of the ambiguities there arises a mixing between `perturbative' effect
and `nonperturbative' effects, rendering it impossible to separate them
in an unique way. Thus, the straightforward Borel resummation defined
on a proper contour  must be augmented by 
nonperturbative effects, which, in general, 
are impossible to calculate.

There can be, however, situations where the Borel resummation of
the perturbation series alone can be useful. For example,
in hadronic tau decay the nonperturbative effects are known to be
small, and so the ambiguities are small, too, to be ignorable. In this case,
roughly speaking, the true amplitude is mostly of perturbative nature,
and can be well described  by the Borel resummation.
Then, the most important thing to do is to describe
the Borel amplitude as accurately as possible in the interval
between the origin and the first IR renormalon using the 
first few perturbative coefficients that are known.

To achieve this purpose, a few techniques were developed. One is to use
conformal transformation to map the UV renormalons far away from the origin, 
which helps accelerate the
convergence of the perturbative expansion of the Borel amplitude.
Another is to use Pad\'e  approximant for the Borel amplitude, either
alone or combined with the conformal mapping.
We introduce in this paper a new technique, which we believe to be 
powerful enough to predict higher order loop corrections, that combines 
the conformal mapping method with a perturbative expansion of the 
Borel amplitude in the neighborhood of the IR renormalon.

Since the ambiguities caused by IR renormalons can be associated with
certain operator condensates, it is possible to expand
the singular part of the Borel amplitude near the renormalon in terms
of the Wilson coefficients and anomalous dimensions 
of the associated operators.
For simplicity, and because we have hadronic tau decay in mind as an 
application of our technique, we shall confine ourselves to the Adler 
function of the current correlators, and the expansion around its first 
IR renormalon caused by gluon condensate.
In Sec.~\ref{sec:bilocal}
we then show that this expansion gives rise to two exact constraints on the
Borel amplitude that need be satisfied at the IR renormalon.
Since one of the constraints involves the 
uncalculated next--to--next--to--leading
order Wilson coefficient of the gluon condensate operator, we have only one
constraint available, which depends only on the calculated 
next--to--leading order Wilson coefficient.
In Sections \ref{sec:conformal} and \ref{sec:prediction}
we then use this constraint to extrapolate, using judiciously
chosen conformal transformations and Pad\'e  approximants 
that involve the unknown ${\cal {O}}(\alpha_s^4)$ 
coefficient of the Adler function,
the perturbative  Borel amplitude in such a way that the constraint be
satisfied at the renormalon. This yields a prediction of the uncalculated
${\cal {O}}(\alpha_s^4)$ coefficient, which we compare with the estimate by  
Kataev and Starshenko \cite{KS} using the method of Stevenson's minimal scale 
dependence, and find it to be consistent with the latter.
We call our method bilocal expansion because the 
constraint  is  derived by using expansion of the Borel amplitude
around the renormalon (around $b\!=\!2$) and the evaluation of the
constraint is carried out by resummations based on the
perturbative expansion of the Borel amplitude (around $b\!=\!0$).

With this prediction of the amplitude up to ${\cal {O}}(\alpha_s^4)$, 
we turn in Sec.~\ref{sec:analysis}
to the hadronic tau decay width without the massive component from
$\triangle S\!\not=\!0$ decays. The width is calculated from
the Adler function by the contour approach in the complex momentum plane.
We use for the Borel transform of the Adler function the ansatz
which explicitly incorporates the structure of the first IR
renormalon, and we perform the Borel integration 
by using an optimal conformal transformation to map away the
effects of the UV renormalons and the higher IR renormalons. 
In Sec.~\ref{sec:alphas},
based on the
stringent experimental results obtained by the ALEPH, OPAL and CLEO
Collaborations, 
we extract with our method of resummation the following
values of the strong coupling parameter:
$\alpha_{\text{s}}(M^2_{\text{z}})
=0.1193 \pm 0.0007_{\rm exp.} \pm 0.0010_{\rm EW+CKM} \pm 0.0009_{\rm meth.}
\pm 0.0003_{\rm evol.}$. 
We compare this result with those of
other variants of our resummation method
and with the present world average,
and subsequently in Sec.~\ref{sec:comparison}
with the results of resummation methods applied previously
by others. Sec.~\ref{sec:summary} contains a summary and
conclusions.

\section{Bilocal expansion}
\label{sec:bilocal}

For definiteness, we shall consider the current-current correlation
function in Euclidean region
\be
\int e^{-i q x}\langle \text{T} J^\mu(x) J^{\nu}(0)^{\dagger}\rangle d^4 x=
-i(q^\mu q^\nu - q^2 g^{\mu\nu}) \Pi(q^2),
\ee
where $J^\mu(x)=\bar{u}\gamma^\mu d(x)$ is the current of up and down
quarks. The canonically normalized massless Adler function 
$D(Q^2)$ is defined by
\be
D(Q^2)\equiv-4\pi^2 Q^2\frac{d}{dQ^2}\Pi(-Q^2) - 1 \ ,
\label{ddef}
\ee
with $Q^2=-q^2 >0$.\footnote{
For normalization convention, see the discussion after
Eq.~(\ref{Acorr}) in the Appendix.}

The Borel transform $\widetilde D(b)$ of the Adler function is defined,
formally, by
\be
D(Q^2)= \frac{1}{\beta_0}
\int^\infty_0 db \,e^{-b/\beta_0 a(Q^2)} \widetilde D(b),
\label{borel-integral}
\ee
where
$a(Q^2)=\alpha_s(Q^2)/\pi$, with $\alpha_s(Q^2)$ being 
the strong coupling constant.
The $\widetilde D(b)$ is analytic around
the origin at $b\!=\!0$, and can be expanded in power series
\be
\widetilde D(b)= 1 + \sum_{n=1}^\infty \frac{d_n}{n!} 
\left(\frac{b}{\beta_0}\right)^n \ ,
\label{borel-def}
\ee
with $d_n$ being the coefficients of the perturbation series for the
Adler function:
\be
D(Q^2)= a(Q^2) \left[ 1 + \sum_{n=1}^{\infty} d_n\,\, a(Q^2)^n \right] \ .
\label{expansion}
\ee
The constant $\beta_0$ is the first coefficient of the QCD $\beta$
function:
\be
\mu^2\frac{d}{d \mu^2} a(\mu^2)\equiv \beta(a(\mu^2))
= - \beta_0 a(\mu^2)^2 [1 + c_1\,\,a(\mu^2) + c_2 \,\,a(\mu^2)^2 +\cdots] \ ,
\label{aRGE}
\ee
where $\mu$ denotes the renormalization scale,
and $c_j\!\equiv\!{\beta}_j/{\beta}_0$ ($j \geq 2$)
parametrize the renormalization scheme.
The Borel transform $\widetilde D(b)$ is known to have singularities:
the UV renormalons on the negative real axis at $b=-n$, and
the IR renormalons on the positive real axis at $b=n+1$ with
$n=1,2,3,\cdots$.
The renormalon resummation of $D(Q^2)$ and of the
hadronic $\tau$ decay width in the large--$\beta_0$
limit has been performed in \cite{BBB,Neubert,LTM}.
While the UV renormalons do not cause any direct problem,
the IR renormalons on the integration contour
cause ambiguities in the Borel integral.
For simplicity, we shall confine ourselves to the first IR renormalon at
$b\!=\!2$. The Borel transform around the singularity can be written
in the form
\be
\widetilde D(b)=\frac{C}{(1-b/2)^{1+\nu}}[
1+{\tilde c}_1 (1-b/2)+ {\tilde c}_2 (1-b/2)^2 +\cdots] +
\left(\text{Analytic part}\right) \ ,
\label{db}
\ee
with $\nu$ given by
\be
\nu= 2 c_1/\beta_0 \approx 1.580
\label{nu}
\ee
for the number of active quark flavors $N_f\!=\!3$. The convergence radius
of the series within the bracket is bounded by the second IR
renormalon at $b=3$, and so the series is expected to be
convergent for $|1-b/2|< 1/2$.

The part analytic at $b\!=\!2$ as well as the exact value of the residue $C$ 
are not known, although the latter can be calculated in perturbation theory
\cite{tlee}.
The coefficients ${\tilde c}_i$ in the expansion 
of the singular part are calculable 
(see Refs. \cite{mueller,beneke} for related discussions),
and depend on the $\beta$ function and the Wilson
coefficients of the gluon condensate operator.

To use this expansion around the renormalon singularity in improving the
Borel resummation, we consider the function $R(b)$
\be
R(b)\equiv (1-b/2)^{1+\nu} \widetilde D(b) \ ,
\label{Rbdef}
\ee
which was introduced in \cite{tlee} in perturbative
calculation of the renormalon residue and also in \cite{soper} to soften the 
renormalon singularity.
Around the singularity, $R(b)$ is given by
\be
R(b)=C [1+ {\tilde c}_1 (1-b/2)+ {\tilde c}_2 (1-b/2)^2 +\cdots] +
(1-b/2)^{1+\nu}\left(\text{Analytic part}\right) \ ,
\label{rb}
\ee
which shows $R(b)$ is singular but bounded at the first IR renormalon.
Should the analytic part vanish, $R(b)$ would be analytic
at the renormalon position, but since there is no reason to expect
this to happen, we should regard $R(b)$ to be singular at $b=2$.
With (\ref{rb}) we now obtain a set of constraints on $R(b)$ 
for $N_f\!=\!3$, and
accordingly on the Borel transform $\widetilde D(b)$, at the singularity
\be
\left. \frac{R'(b)}{R(b)}\right|_{b=2}=-\frac{{\tilde c}_1}{2} \ ,
\hspace{0.5in}
\left. \frac{R''(b)}{R(b)}\right|_{b=2}=\frac{ {\tilde c}_2}{2} \ .
\label{constraints}
\ee
In the next Section we will exploit one of these equations to constrain
the functional behavior of the Borel transform in the interval
between the origin and the first IR renormalon singularity.

We now turn to the calculation of the coefficients 
${\tilde c}_1, {\tilde c}_2$.
Because of the singularity, $\widetilde D(b)$ has a branch cut beginning at
$b=2$, and consequently, $D(Q^2)$ obtains an imaginary
part from the Borel integral:
\be
\text{Im}[D(Q^2)] \propto \pm a(Q^2)^{-\nu} e^{-2/\beta_0 a(Q^2)}
\left[1+ \frac{1}{2}{\tilde c}_1 \nu \beta_0\,\,a(Q^2)+
\frac{1}{4}{\tilde c}_2\nu (\nu-1)\beta_0^2
\,\,a(Q^2)^2 +O(a^3)\right],
\label{imag1}
\ee
which is obtained by plugging (\ref{db}) into (\ref{borel-integral}).
The sign of the imaginary part depends on whether the contour along the
positive real axis is on the upper or the lower half plane.
Because $D(Q^2)$ must be real, this imaginary part should be canceled
by something else. It has been suggested in \cite{david} that this
imaginary part is canceled by the
imaginary part arising from the ambiguity in defining  renormalized
gluon condensate  in the
Operator Product Expansion (OPE) of $D(Q^2)$
\be
D(Q^2)= {\cal C}_0(a(Q^2)) +{\cal C}_4(a(Q^2))\frac{\langle
O_4\rangle}{Q^4} +\left(\text{Higher dimension terms}\right) \ ,
\label{OPE}
\ee
where $\langle O_4 \rangle$ is the scale--invariant
matrix element (gluon condensate)
of the anomalous--dimension free,
dimension--four gluon operator
\be
\langle O_4 \rangle =  \langle \frac{\beta(a)}{a}
G^a_{\mu\nu}G^{a\mu\nu} \rangle \ ,
\label{O4}
\ee
with $G^a_{\mu\nu}$ denoting the gluon field strength tensor.

The Wilson coefficient ${\cal C}_0$ for the unit operator
has the perturbative expansion given in (\ref{expansion}) while the
coefficient
${\cal C}_4$ for the gluon condensate operator is known to the
next--to--leading order (NLO) in the $\overline {\rm MS}$ renormalization
scheme \cite{wilson}
\be
{\cal C}_4(a(Q^2)) =-\frac{2\pi^2}{3\beta_0}\left[1 +w_1\,\, a(Q^2)
+w_2 \,\,a(Q^2)^2 +O(a^3)\right] \ ,
\label{C4}
\ee 
with 
\be
w_1=\frac{7}{6}-c_1  \ \left(= - \frac{11}{18} \ {\rm for} \
N_f\!=\!3 \right) \ .
\label{w1}
\ee
The next--to--next--to--leading order (NNLO) 
coefficient $w_2$  is not known yet.

Because the gluon condensate as well as its ambiguity should satisfy the
homogeneous renormalization group (RG) equation, 
the ambiguous, imaginary part
from the gluon condensate  can be written as
\be
\text{Im} [D(Q^2)_{\text{con.}}]=
\pm \,\,{\cal C}_4 (a(Q^2))\,\, \frac{\Lambda^4}{Q^4} \ ,
\label{imag2}
\ee
with $\Lambda$ being a RG--invariant and $Q$--independent
constant. Therefore,
\bear
\frac{{\Lambda}^4}{Q^4}  &\propto&
\exp \left[ - 2 \int^{a(Q^2)} \frac{dx}{\beta(x)} \right] \nonumber \\
 &\propto&  a(Q^2)^{-\nu} e^{-2/\beta_0 a(Q^2)}  \left[1 +v_1 a(Q^2)
+v_2 a(Q^2)^2
+O(a^3)\right] \ ,
\label{prop}
\eear
where the proportionality constants are $Q$--independent, and
$v_j$'s are obtained by expanding $1/{\beta}(x)$ in powers of $x$
\bear
v_1&=&\frac{2}{\beta_0} ( -c_2 + c_1^2 ) \ ,
\nonumber \\
v_2&=& \frac{1}{2}v_1^2 +\frac{1}{\beta_0} ( - c_3 + 2 c_1 c_2 - c_1^3) \ .
\label{v12}
\eear
Thus, 
\begin{eqnarray}
\text{Im}[ D(Q^2)_{\text{con.}}] & \propto & 
\pm\,a(Q^2)^{-\nu} e^{-2/\beta_0 a(Q^2)}
\nonumber\\
& &\times 
\left[ 
1+(v_1\!+\!w_1)\,\, a(Q^2)+ (v_2\!+\!v_1 w_1\!+\!w_2) a(Q^2)^2 +O(a^3)
\right] \ .
\label{imag3}
\end{eqnarray}

Because the imaginary parts in (\ref{imag1}) and (\ref{imag3}) should
cancel each other, we have
\bear
{\tilde c}_1&=& \frac{2}{\nu\beta_0} (v_1+w_1)  \ 
\left( \approx -0.9990 \ {\rm for} \ N_f\!=\!3 \right) \ ,
\label{coeff1}
\\
{\tilde c}_2&=&\frac{4}{\nu(\nu-1)\beta_0^2}(v_2+v_1 w_1 +w_2) \ .
\label{coeff2}
\eear

\section{An optimal conformal mapping}
\label{sec:conformal}

To impose the constraints (\ref{constraints}) on the Borel transform
defined in series form (\ref{borel-def}), $\widetilde D(b)$ needs
be analytically continued beyond its convergence radius $|b|=1$
which is set by the first UV renormalon. This cumbersome,
analytic continuation however can be conveniently avoided by using a
conformal mapping that pushes the UV renormalons away from the origin
while mapping the first IR renormalon to be the closest singularity
to the origin.
Since, in practice, only the first few coefficients are known,
choosing an optimal mapping can help accelerate convergence
of the series (\ref{borel-def}). Even though several conformal mappings,
being optimal or not, were discussed in the literature 
\cite{mueller1,tlee,caprini} we introduce a new
mapping which is especially well--suited for our purpose.

Our criterion for an optimal mapping is simple; with an optimal mapping
\be
w=w(b)
\label{om}
\ee
the function $R(b(w))$ should be as smooth as possible within
the disk $|w|\le w_0$, where $w_0=|w(b=2)|$, so that
 $R(b(w))$ within the radius of convergence can be well approximated
by the first terms of its perturbation series in $w$
\be
R(b(w))=\sum_{n=0}^\infty r_n w^n.
\label{r-series}
\label{Rbexp}
\ee

\begin{figure}[ht]
 \centering\epsfig{file=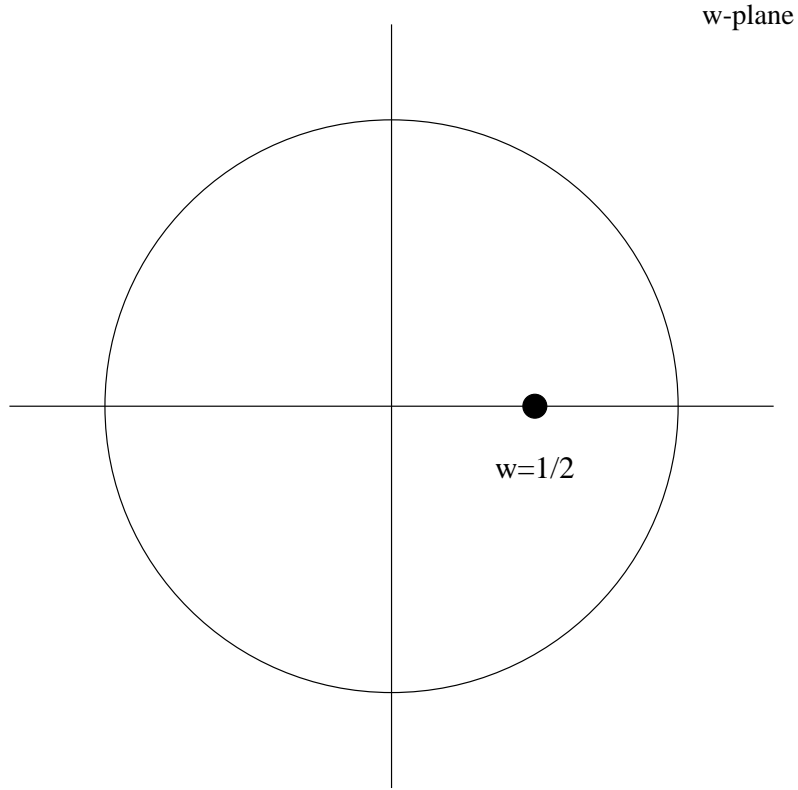}
\vspace{0.3cm}
\caption{\footnotesize
The conformal mapping (\ref{mapping}) maps the first IR renormalon
to $w=1/2$, and all other renormalons to the unit circle.}
\label{fig1}
\end{figure}

With this criterion our strategy for an optimal mapping is to
send all the renormalon singularities save the unavoidable first
IR renormalon as far away as possible from the origin.
As a candidate for an optimal mapping we propose
\be
w=\frac{\sqrt{1+b}-\sqrt{1-b/3}}{\sqrt{1+b}+\sqrt{1-b/3}} \ ,
\label{mapping}
\ee
which is obtained by combining the mapping \cite{tlee}
\be
z=\frac{b}{1+b} \ ,
\ee
which  sends all the UV renormalons to the positive real axis,
with the  mapping \cite{mueller1}
\be
w=\frac{1-\sqrt{1-z/z_0}}{1+\sqrt{1-z/z_0}} \ ,
\ee
where $z_0\equiv z(b=3)=3/4$, that sends all renormalon singularities 
except for the first IR renormalon to the unit circle.
With the conformal mapping (\ref{mapping}) 
the first IR renormalon is mapped
to $w=1/2$ while all other renormalons are mapped to the unit circle
(see ~Fig.~\ref{fig1}).
Since we are especially interested in the functional behavior of
$R(b(w))$ within the radius of 
convergence $w_0=1/2$, we expect the mapping
is well-suited for our purpose because the divergence by the
renormalon singularities  are suppressed due to 
their relatively large distance to the origin.

Now on the $w$-plane the first of the constraints (\ref{constraints})
becomes
\be
\left. \left[\frac{d R(b(w))}{d w} + \frac{{\tilde c}_1}{2}  
\frac{d b}{dw} R(b(w)) \right]\right|_{w=\frac{1}{2}}=0 \ .
\label{constraint2}
\ee
In the next Section we will impose this constraint on
the truncated perturbation series (TPS) of (\ref{r-series}) 
to obtain a higher order
correction of the Adler function. By noticing
the constraint (\ref{constraint2})
is set up at the first IR renormalon, which is exactly at the
radius of convergence of the series (\ref{r-series}),
one may question the validity
of applying the constraint directly on the TPS.
However, it should be emphasized that the 
series (\ref{r-series}) is convergent
at the renormalon singularity $w=1/2$ 
because $R(b(w))$, even though singular
there, is bounded. Therefore,  the constraint can be imposed on 
the perturbation series.

\section{Prediction for the NNNLO coefficient of the Adler function}
\label{sec:prediction}

The NLO and NNLO coefficients $d_1$ and $d_2$ of the
expansion of the canonical Adler function (\ref{expansion})
have been calculated exactly  in the $\overline {\rm MS}$
scheme in \cite{coeffs1,coeffs2}: 
$d_1\!=\! 1.6398$, $d_2\!=\!6.3710$ (at  $N_f\!=\!3$). 
The Borel transform ${\widetilde D}(b)$ (\ref{borel-def})
and the function $R(b)$ (\ref{Rbdef}) are thus also known up to NNLO
in $b$. Upon subsequently applying the conformal transformation 
(\ref{mapping}) to $R(b)$, and expanding in $w$, we obtain
the power expansion of $R(b(w))$ (\ref{Rbexp}) up to NNLO in $w$.
On the other hand, if we assumed that the ${\rm N}^3 {\rm LO}$
coefficient $d_3$  were known, we would obtain the power expansion
(\ref{Rbexp}) up to ${\rm N}^3 {\rm LO}$
\be
R(b(w)) = 1. - 1.68394 w + 0.104 w^2  + (-9.64395 + 0.395062 d_3) w^3 
 +  O(w^4) \ .
\label{RbwTPS}
\ee
The corresponding derivative $dR/dw$ would then be known up to NNLO
($\sim\!w^2$). If we apply to the constraint (\ref{constraint2})
the above  ${\rm N}^3 {\rm LO}$ TPS of $R$ and the corresponding
NNLO TPS for $dR/dw$, we obtain $d_3 \approx 34$.
This prediction, however, is not sufficiently precise, 
because, as mentioned before, the point $w\!=\!1/2$ is
at the border of the convergence disk of $R(b(w))$ and
we are dealing with strongly truncated series. Therefore,
we apply at this stage yet another efficient mechanism of analytic
continuation which would bring us beyond the $w\!=\!1/2$ circle
-- Pad\'e approximants (PA's)\footnote{
The authors of Ref.~\cite{JS} showed that 
combining the conformal transformations with the PA--type
of resummations can lead to significantly improved results,
at least when a sufficient number of terms in the power
expansion are known.}
that are either diagonal or 
near--diagonal \cite{Pade}. 
To the ${\rm N}^3 {\rm LO}$ TPS (\ref{RbwTPS}) 
of $R(b(w))$ we can then either
apply the $[1/2]$, $[2/1]$, or $[1/1]$ PA, and to the NNLO TPS
of $dR/dw$ the $[1/1]$ PA. Then the constraint (\ref{constraint2})
predicts the values $d_3 \approx 24.7$--$24.8$, virtually
independent of the three PA--choices for $R(b(w))$.\footnote{
In the procedure, we further require that $[1/1]$ PA of $dR/dw$
not possess clearly unphysical poles (i.e., poles well below $w=0.5$).}
Another practical approach is to construct, at a given fixed
$d_3^{(0)}$, the NNLO TPS of $d \ln R/dw$ and thus the
PA $[1/1]$ of $d \ln R/dw$. Employing this PA in the
constraint (\ref{constraint2}) leads to the
prediction $d_3^{(0)} \approx 30.4$. In the latter approach,
however, higher order PA's ($[2/1]$, $[1/2]$) cannot be employed.

As a cross--check, we carried out the same procedure,
but with a different conformal transformation
\be
w=\frac{\sqrt{1+b}-\sqrt{1-b/4}}{\sqrt{1+b}+\sqrt{1-b/4}} \ .
\label{mapping4}
\ee
This mapping also removes all the UV renormalons 
to the unit circle, as well as all the IR renormalons
except for the first ($b\!=\!2$) and the second ($b\!=\!3$) one:
$w(b\!=\!2) \approx 0.42$,
$w(b\!=\!3)=0.6$. This mapping apparently suppresses even
more strongly than (\ref{mapping}) the UV renormalon contributions,
but probably less strongly the next--to--leading
IR renormalon ($b\!=\!3$) contributions. The predictions
are in this case $d_3 \approx 24.3$--$24.5$, in good
agreement with the aforementioned predictions.
The use of the PA $[1/1]$ of $d \ln R/dw$ predicts in this case
$d_3^{(0)} \approx 30.3$.


We further note another interesting feature of the
expansion (\ref{RbwTPS}). Looking at the first three terms that are known,
it appears reasonable to expect that
the ${\rm N}^3 {\rm LO}$ coefficient $r_3$ at $w^3$ is not very large,
say, $|r_3| < 2$. Varying $r_3$ between $-2$ and $2$
results in the variation of $d_3$ between $19.3$ and
$29.5$, i.e., only about $20 \%$ around the  
value $d_3 \approx 24.7$.
Thus, the
predictions of the described method, using the
conformal transformation (\ref{mapping}), are
remarkably robust under the variation of the
${\rm N}^3 {\rm LO}$ coefficient of $R(b(w))$.
Similar robustness is observed when the
conformal transformation (\ref{mapping4}) is used
instead of (\ref{mapping}).

If we applied to the relation (\ref{constraint2}) the
PA's, but no conformal transformation, the predictions
would vary more strongly ($d_3 \approx 26.$--$33.$)
with the various choices for PA's of $R(b)$.
Furthermore, this method does not possess the 
`robustness' under the variation of the
${\rm N}^3 {\rm LO}$ coefficient of $R(b)$.

Thus, our prediction of the ${\rm N}^3 {\rm LO}$
coefficient $d_3$, in the $\overline {\rm MS}$ scheme,
of the  Adler function $D(Q^2)$ is
\be
d_3 \approx 25. \pm 5.  \,\,\,\,(\text{at $N_f=3$}) \,
\label{d3pr}
\ee
 based on the
simultaneous use of relation (\ref{constraint2}), 
the conformal mapping (\ref{mapping}), and the
Pad\'e approximants. However, a larger uncertainty
($\pm 10.$) of the predicted values of $d_3^{(0)}$
cannot be excluded, and we will use these more
conservative uncertainty estimates in the next Sections
[see (\ref{d3o0var})].

Our predictions can be compared, for example, with those of
Ref.~\cite{KS}. They used the method of effective charge (ECH)
\cite{ECH,KKP,Gupta} and the TPS principle of minimal
sensitivity (PMS) \cite{PMS,PMS2} for the NNLO TPS of the
 Adler function $D(Q^2)$. The obtained approximants
were then re--expanded back in powers of 
$a_0\!\equiv\!a(Q^2;{\overline {\rm MS}})$ up to $\sim\!a_0^4$,
under the assumption 
$c_3^{\rm ECH}\!-\!c_3^{{\overline {\rm MS}}} \approx 0$
and
$c_3^{\rm PMS}\!-\!c_3^{{\overline {\rm MS}}} \approx 0$.
The resulting prediction was $d_3 \approx 27.5$, 
which is consistent with our prediction (\ref{d3pr}).

\section{Analysis of the hadronic tau decay}
\label{sec:analysis}

In this Section we will apply elements of the previous
Sections to the numerical study of the $\tau$ inclusive hadronic
 decay  ratio
\begin{equation}
R_{\tau} \equiv \frac{ 
\Gamma (\tau^- \to \nu_{\tau} {\rm hadrons} (\gamma) )}
{ \Gamma (\tau^- \to \nu_{\tau} e^- {\overline {\nu}_e} (\gamma))} \ .
\label{Rtaudef}
\end{equation}
Here, $(\gamma)$ represent possible additional photons,
or lepton pairs.
This inclusive decay ratio has been extensively studied in the
literature, theoretically and numerically
\cite{rtau0,rtau1,rtau2,rtau3,rtau4,rtau6,rtau7,BNP}. 
The ratio can be expressed, via the application of a
variant of the optical theorem, with the two-point
correlation functions of the vector (V) and axial--vector (A)
currents, or equivalently, with the Adler functions
$D^{\rm L+T}(Q^2)$ and $D^{\rm L}(Q^2)$ --
we refer to the Appendix for some details. 
The theoretical/numerical resummation methods
for evaluation of QCD observables are most
efficient in the limit of massless quarks.
When excluding hadrons with $s$ (strange) quarks,
and approximating the $u$ and $d$ quarks to be massless,
the expression can be written as a contour integral (\ref{rtmud2=0})
in the complex momentum plane \cite{rtau3,rtau4,rtau6,rtau7,BNP}
\begin{eqnarray}
r_{\tau} & \equiv & 
r_{\tau}^{{\rm V\!+\!A}}(\triangle S\!=\!0; m_{u,d}\!=\!0) 
 \equiv 
\frac{ R_{\tau}^{{\rm V\!+\!A}}(\triangle S\!=\!0; m_{u,d}\!=\!0) }
{ 3 |V_{ud}|^2 (1 + {\delta}_{{\rm EW}}) } -
(1 + \delta_{\rm EW}^{\prime} )
\label{rtdef}
\\
&=& \frac{1}{2 \pi} \int_{-\pi}^{\pi} dy \left( 1 + e^{{\rm i} y} \right)^3
\left( 1 - e^{{\rm i} y} \right) 
D(Q^2\!\equiv\!-s\!=\!m_{\tau}^2 e^{{\rm i}y} ) \ ,
\label{contint}
\end{eqnarray}
where the (minimal Standard Model) electroweak correction (EW) factors 
$\delta_{\rm EW}$ and $\delta^{\prime}_{\rm EW}$
have been calculated in \cite{MS,BL},
and $D(Q^2)$ is the massless
canonical Adler function (\ref{ddef}), (\ref{expansion}).
The superscript ${\rm V\!+\!A}$ in the above formulas
emphasizes the fact that the quantities are inclusive in the sense
of including the vector and axial--vector hadronic currents.

The experimental value of the
observable $R_{\tau}^{{\rm V\!+\!A}}(\triangle S\!=\!0)$
can be extracted from the values of the leptonic
branching ratios
$B_e \equiv B( \tau^- \to e^- {\overline \nu}_e \nu_{\tau})$ and
$B_{\mu} \equiv B( \tau^- \to \mu^- {\overline \nu}_{\mu} \nu_{\tau})$,
as obtained from the constrained fit derived from a set of basis
modes \cite{PDG2000} (see also \cite{ALEPH1}). The basis modes
form an exlusive set of leptonic and hadronic decays
whose branching ratios are normalized so that their sum
is exactly one. 
The set of basis modes does not
include the decays with photons in the final state,
i.e., the right--hand side of (\ref{Rtaudef}) is
for them without $(\gamma)$. 
The only leptonic branching ratios in
the set of basis modes are $B_e$ and $B_{\mu}$.
Therefore, $R_{\tau} = (1 - B_e - B_{\mu})/B_e$.
The present values of $B_e$ and $B_{\mu}$, 
as determined from the constrained fit
\cite{PDG2000}, based on the high precision measurements of the
basis modes of the $\tau^-$ decay by the ALEPH,
OPAL, and CLEO Collaborations
\cite{ALEPH1,ALEPH2,Davier,OPAL,CLEO}, are:
$B_e = (17.83 \pm 0.06) \times 10^{-2}$,
$B_{\mu} = (17.37 \pm 0.07) \times 10^{-2}$.
The updated value of the strangeness--changing ratio is
\cite{Davier2}
$R_{\tau}(\triangle S\!\not=\!0) = 0.1630 \pm 0.0057$.
This implies
\begin{eqnarray}
R_{\tau}^{{\rm V\!+\!A}}(\triangle S\!=\!0) &=&
\frac{(1 - B_e - B_{\mu} )}{B_e} -
R_{\tau}(\triangle S \!\not=\!0)
\label{Rtaudef2}
\\
& = & 3.4713 \pm 0.0171 \ .
\label{Rtauexp}
\end{eqnarray}
The canonical $\tau^-$ decay ratio (\ref{rtdef}), but 
at the moment still without the massless quark condition $m_{u,d} \to 0$,
i.e., the reduced decay ratio (\ref{rtgen}) in the Appendix,
can then be obtained from the experimental values (\ref{Rtauexp})
by inserting the known values of the electroweak correction
parameters $\delta_{\rm EW}$ and $\delta^{\prime}_{\rm EW}$
and of the Cabibbo--Kobayashi--Maskawa (CKM) matrix
element $|V_{ud}|$. Here we have to deal
with additional uncertainties. 

The main EW correction parameter has the value 
$\delta_{\rm EW} = 0.0194 \pm 0.0050$ \cite{MS},
while the residual correction parameter
is $\delta^{\prime}_{\rm EW} = 0.0010$ \cite{BL}.
In calculating $\delta_{\rm EW}$, the additional contributions
from low scales ($<\!m_{\tau}$), dependent on the hadronic 
structure, although not enhanced by large logarithms,
cannot be calculated and were estimated \cite{MS} to lead
to the significant uncertainties $\pm 0.0050$.

The values of $|V_{ud}|$ from the (SM) unitarity
constraint fit are $0.9749 \pm 0.0008$ \cite{PDG2000}.
On the other hand, the values extracted from the
decays of mirror nuclei lead to lower values
$|V_{ud}| = 0.9740 \pm 0.0010$. This extraction is,
however, frought with theoretical uncertainties
(see \cite{PDG2000} for further References).
Further, the values extracted from
neutron decays $|V_{ud}| = 0.9728 \pm 0.0012$
(\cite{PDG2000} and References therein)
are even lower, but
appear to have smaller theoretical uncertainties.
For all these reasons, we will adopt the value range
\begin{equation}
|V_{ud}| = 0.9749 \pm 0.0021 \ ,
\label{Vud}
\end{equation}
where the central value is the one from the
unitarity constrained fit, but the uncertainty
has been increased so that the values now
include all the values from the decays of mirror nuclei
and the upper half of the interval of values from 
neutron decays.

This now allows us to extract the values of the
canonical $\tau^-$ decay ratio (\ref{rtgen})
\begin{eqnarray}
r_{\tau}^{{\rm V\!+\!A}}(\triangle S\!=\!0) & \equiv & 
\frac{ R_{\tau}^{{\rm V\!+\!A}}(\triangle S\!=\!0) }
{ 3 |V_{ud}|^2 (1 + {\delta}_{{\rm EW}}) } -
(1 + \delta_{\rm EW}^{\prime}) 
\nonumber\\
&=& 0.1933 \pm 0.0059_{\rm exp.}
\pm 0.0059_{\rm EW} \pm 0.0051_{\rm CKM} \ ,
\label{rtauexpmud}
\end{eqnarray}
where the uncertainty $\pm 0.0059_{\rm EW}$ originates
from the aforementioned $\pm 0.0050$ uncertainty in
$\delta_{\rm EW}$, and $\pm 0.0051_{\rm CKM}$
from the $\pm 0.0021$ uncertainty in $|V_{ud}|$ (\ref{Vud}).

The QCD observable (\ref{rtauexpmud}), as defined,
has the non-QCD effects factored out. However, it still
contains the problematic, though small, quark mass effects
($m_{u,d} \not= 0$). In the Appendix, we calculated the
numerical strength of the
quark mass contributions (\ref{drtmud1})--(\ref{drtmud2}).
Subtracting these effects as in (\ref{rtmud1=0}), we end up with
the following values for the massless QCD observable (\ref{rtdef})
\begin{eqnarray}
r_{\tau} &\equiv& r_{\tau}^{{\rm V\!+\!A}}(\triangle S\!=\!0; m_{u,d}\!=\!0)
\nonumber\\
&=& 0.1960 \pm 0.0059_{\rm exp.} \pm 0.0059_{\rm EW} \pm 0.0051_{\rm CKM}
\label{rtauexp}
\\
&=& 0.1960 \pm 0.0098 \ .
\label{rtauexps}
\end{eqnarray}
In (\ref{rtauexp}) we neglected the small contributions
$\sim\!0.0001$ from the corrections of the type 
$\sim\!m^2_{u,d}/m^2_{\tau}$ to (\ref{drtmud2}).
In (\ref{rtauexps}), the
three uncertainties of (\ref{rtauexp}) were added in quadrature.

The values (\ref{rtauexp}) will be the starting
point for our massless QCD resummation analyses of the hadronic
$\tau$ decay.
The experimental uncertainty (\ref{rtauexp}) in the 
massless QCD observable $r_{\tau}$ is $3 \%$, 
representing a high experimental precision
when compared to many other QCD observables. This fact can
be regarded at present as our main motivation to
investigate theoretically and numerically this observable.
Unfortunately, as we can see from (\ref{rtauexp})--(\ref{rtauexps}),
the total precision is worse ($5 \%$), due
to the present uncertainties in the values of the
electroweak corrections and of $|V_{ud}|$. 

By adjusting the numerical (resummed) predictions for $r_{\tau}$
to the experimental ones (\ref{rtauexp}), our main goal
will be to predict the QCD coupling parameter
$\alpha_s(m^2_{\tau})$ with the high precision, i.e.,
with the resummation method uncertainty 
$(\delta \alpha_s)_{\rm meth.}$
of the prediction being comparable to, or smaller than, the
experimental uncertainty $(\delta \alpha_s)_{\rm exp.}$
stemming from $({\delta} r_{\tau})_{\rm exp.}\!=\!0.0059$
(\ref{rtauexp}). The starting point for our resummation
method will be the contour integral representation
(\ref{contint}) of $r_{\tau}$ in terms of the
massless Adler function $D(Q^2)$.

\begin{figure}[ht]
 \centering\epsfig{file=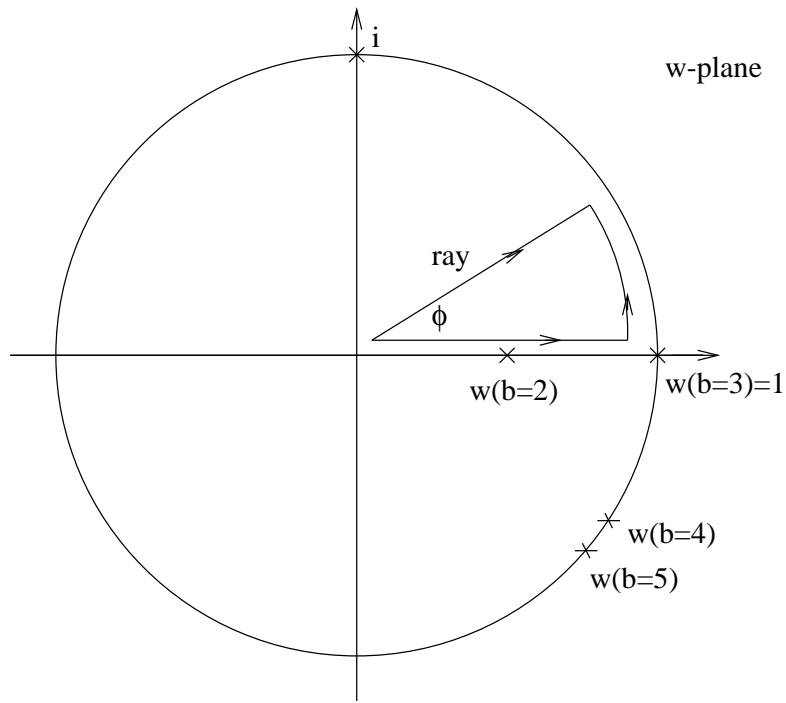}
\vspace{0.3cm}
\caption{\footnotesize
Integration along the ray
$w = x \exp({\rm i} \phi)$ ($0 < x < 1$, $\phi$ fixed)
gives the same result as the integration parallel to the
positive real axis ($0 < w < 1$) and arc
$w = \exp({\rm i} {\phi}^{\prime})$ ($0 < {\phi}^{\prime} < \phi$).}
\label{fig2}
\end{figure}

As in the previous Sections, we express $D(Q^2)$
as the Borel integral of its Borel transform 
$\widetilde D(b)=R(b)/(1-b/2)^{1+\nu}$, with 
the correct first IR renormalon singularity
explicitly enforced in the ansatz
\begin{equation}
D(Q^2) = \frac{1}{\beta_0} \text{Re}\left[ \int_{0+i\varepsilon}
^{\infty+i\varepsilon} db\,\, e^{-b/\beta_0 a(\xi^2 Q^2)} 
\frac{R(b; \xi^2)}{(1 - b/2)^{1+\nu}}\right] \ ,
\label{BTd}
\end{equation}
where the integration contour
is chosen  to be on the upper half plane to avoid the singularity at
$b\!=\!2$ (Cauchy principal value prescription). 
By explicitly enforcing the renormalon singularity, the Borel
transform around the singularity can be more accurately described, and
also the validity of the perturbative Borel transform can be extended
beyond the first IR renormalon.
The Borel transform  ${\widetilde D}(b)$ as well as $R(b)$ depend
on the renormalization scheme, and on the renormalization scale
parameter $\xi^2\!=\!\mu^2/Q^2$ through
the $\xi^2$--dependence of the perturbative coefficients $d_n$
in (\ref{expansion}) when the running coupling $a(Q^2)$ is replaced by
$a(\xi^2 Q^2)$. While we choose $\overline{\rm MS}$ scheme throughout this
paper, the renormalization scale parameter $\xi^2$ will be kept 
arbitrary for the time being. When inserting (\ref{BTd}) in (\ref{contint}),
and exchanging the order of integrations, we obtain
\begin{eqnarray}
r_{\tau} & = &
\frac{1}{2 \pi \beta_0} \text{Re}\left[ \int_{0+i\varepsilon}^
{\infty+i\varepsilon} db 
\frac{R(b; \xi^2)}{(1 - b/2)^{1+\nu}}
\times \right.
\nonumber\\
&&
\left. \int_{-\pi}^{\pi} dy \left( 1 + e^{{\rm i} y} \right)^3
\left( 1 - e^{{\rm i} y} \right)
e^{
- b/\beta_0 a \left(
\xi^2 m_{\tau}^2 \exp({\rm i} y)\right) } \right] \ .
\label{rt2}
\end{eqnarray}
Since the integrand is exponentially suppressed at large $b$,
it is convenient and reasonable to integrate over the Borel
variable $b$ just to a certain value $b_{\rm max}$ lying
beyond the first IR renormalon. The contribution from the 
region beyond the first IR renormalon is expected to be smaller
or comparable to the nonperturbative effect by the
the gluon condensate, which is known to be small \cite{BNP}.
If we know the  perturbation series  of
the Adler function $D(Q^2)$ up to
${\rm N}^3 {\rm LO}$  then
we know automatically also $R(b; \xi^2)$
up to ${\rm N}^3 {\rm LO}$, i.e., including the term $\sim\!b^3$.
Further, $R(b; \xi^2)$ has no singularities on the
positive axis for $b < 2$, and only a soft singularity at $b=2$,
but it has some UV renormalons on the
negative axis rather close to the origin: $b = -1, -2$.
These UV renormalons make the power expansion of $R(b; \xi^2)$
in powers of $b$ divergent for $|b|\geq 1$, which signals that
the use of the (${\rm N}^3 {\rm LO}$) TPS in powers of $b$ for
$R(b; \xi^2)$ in (\ref{rt2}) may run into
serious trouble already at $b \geq +1$. An efficient solution
to this problem was already constructed in Section \ref{sec:conformal},
in the form of an optimal conformal transformation
$b\!=\!b(w)$ (\ref{mapping}), which pushes all the UV renormalons
(and all the higher IR renormalons at $b \geq 3$)
onto a unit circle in the plane of the new variable $w$.
The first IR renormalon at $b\!=\!2$ now
corresponds to $w\!=\!1/2$, i.e., within the unit circle.
Then, the expansion $R(b(w); \xi^2)$
in powers of $w$ represents a convergent series for $w \leq 1/2$,
i.e., for the corresponding  $b(w) \leq 2$. Thus, the use of the
corresponding ${\rm N}^3 {\rm LO}$ TPS of $R(b(w); \xi^2)$, 
which is also explicitly known,
will have much better chances to describe reasonably well
the true $R(b(w); \xi^2)$ within the interval between the origin and
the first IR renormalon. Therefore, the double integral
(\ref{rt2}) will be rewritten in terms of the variable $w$
\begin{eqnarray}
\lefteqn{
r_{\tau}  \approx
\frac{1}{2 \pi \beta_0} \text{Re}\left[ 
\int_{0}^{w_{\rm max}} dw 
\frac{d b(w)}{d w} 
\frac{R(b(w); \xi^2)}
{(1 - b(w)/2)^{1+\nu}}
\times \right.} 
\nonumber\\
&&
\left. \int_{-\pi}^{\pi} dy \left( 1 + e^{{\rm i} y} \right)^3
\left( 1 - e^{{\rm i} y} \right)
e^{- b(w)/\beta_0 a \left(
\xi^2 m_{\tau}^2 \exp({\rm i} y) \right) } \right]
\label{rt3a}
\\
&=&
\frac{3}{2 \pi \beta_0} {\rm Re} \left[  e^{{\rm i} \phi} 
\int_0^1 dx 
(1 - w^2) (1 - w + w^2)^{\nu-1}
\frac{R(b(w); \xi^2)}
{ (1/2 - w)^{1+\nu} (2 - w)^{1+\nu} } \times \right.
\nonumber\\
&& 
\left. \left.
\int_{-\pi}^{\pi} dy \left( 1 + e^{{\rm i} y} \right)^3
\left( 1 - e^{{\rm i} y} \right)
e^{- b(w)/\beta_0 a \left(
\xi^2 m_{\tau}^2 \exp({\rm i} y)\right) } 
\right|_{w=x {\rm e}^{ {\rm i} \phi}}
\right] \ ,
\label{rt3}
\end{eqnarray}
where we can choose in (\ref{rt3a}) $w_{\rm max} \gg 1/2$,
corresponding to $b_{\rm max} \gg 2$. In practice, we
can go in the $d w$--integration in (\ref{rt3a}) 
beyond $w\!=\!1.$, where the $w$--contour follows then the unit 
circle arc into the first quadrant -- for example up to a complex 
$w_{\rm max} = \exp( {\rm i} \phi)$ with 
$0<\phi < \phi_\infty$, where $w(b\!=\!\infty) = 
\exp(  {\rm i} {\phi}_\infty)$,
$\phi_\infty =\pi/3$.
The fact that in this way we reach the $b \approx 3$ region,
where the true $R(b)$ has an IR renormalon,
and even go beyond it, does not change the
result of (\ref{rt3a}) in practice. This is so because
the contributions from the arc
$|w|\!=\!1$ (corresponding to $b \stackrel{>}{\sim} 3$)
turn out to be extremely suppressed in (\ref{rt3a})
(see also footnote \ref{wmax} of the next Section).
This integration can be implemented in
practice most easily, if we follow the ray
$w\!=\!x \exp( {\rm i} {\phi})$, with $x$ from $0$ to $1$
(see Fig.~\ref{fig2}), because the integration over the
corresponding closed contour yields zero since no singularities are
enclosed. This practical `ray'--integral
implementation is denoted in (\ref{rt3}).

The first two coefficients $d_1$ and $d_2$ of the
expansion of the  Adler function (\ref{expansion}),
which determine the expansions of $\widetilde D (b)$ and
$R(b)$ up to NNLO,
have been calculated exactly in the literature
\cite{coeffs1,coeffs2}. For the choice
$\mu^2\!=\!Q^2$ and in the ${\overline {\rm MS}}$ scheme,
with $N_f\!=\!3$, they are: $d_1^{(0)}\!=\! 1.6398$,
$d_2^{(0)}\!=\!6.3710$. In the previous Section,
the arguments were presented suggesting the value of the
${\rm N}^3{\rm LO}$ coefficient: $d_3^{(0)}\approx 25.$
When the renormalization scale $\mu^2\!=\!\xi^2 Q^2$ is changed
($\xi^2 \not= 1$), these coefficients change accordingly:
\begin{eqnarray}
d_1 &=& d_1^{(0)} + \beta_0 \ln \xi^2 \ ,
\label{d1xi2}
\\
d_2 &=& d_2^{(0)} + 2 \beta_0 \ln \xi^2 d_1^{(0)} + \beta_1 \ln \xi^2
+ (\beta_0 \ln \xi^2)^2 \ ,
\label{d2xi2}
\\
d_3 &=& d_3^{(0)} + 
3 (d_1 d_2\!-\!d_1^{(0)} d_2^{(0)}) - 2 (d_1^3\!-\!d_1^{(0)3})
-(c_1/2) (d_1^2\!-\!d_1^{(0)2}) + 
c_2 (d_1\!-\!d_1^{(0)}) \ .
\label{d3xi2}
\end{eqnarray}
These relations follow from the expressions for the
renormalization scheme and scale--invariants $\rho_1$, $\rho_2$, $\rho_3$, 
as given, e.g., in \cite{PMS}.
As an example, at $\xi^2\!=\!2$ they imply: $d_1\!=\!3.1994$,
$d_2\!=\!16.6908$, $d_3\!=\!97.4436$.
The corresponding ${\rm N}^3 {\rm LO}$ 
Borel transform is:
\begin{eqnarray}
R(b(w); \xi^2) 
&=& 1 - 1.68394 w + 0.104 w^2  + 0.232591 w^3 \quad (\xi^2\!=\!1) \ ,
\label{tbxi2=1}
\\
&=& 1 + 0.395499 w + 3.30834 w^2  + 5.13735 w^3 \quad (\xi^2\!=\!2) \ .
\label{tbxi2=2}
\end{eqnarray}
The apparently quite strong $\xi^2$--dependence of the
Borel transform function
$R(b(w); \xi^2)$
in (\ref{rt3})
is combined with the strong $\xi^2$--dependence of the
coupling parameter $a(\xi^2 Q^2)$ in
the exponent (\ref{rt3}) in such a way that the entire
double integral is $\xi^2$--independent. However, since
we know just the first few terms
of $R(b(w); \xi^2)$,
the $\xi^2$--dependence of (\ref{rt3}) will appear.
If the method is good, this dependence should be weak,
at least locally in a renormalization scale region $\xi^2\!\sim\!1$.
Further, there should be
some  dependence on the choice of the renormalization
scheme,
but the scheme dependence is in general weaker than
the $\xi^2$-- dependence, and we choose $\overline{\rm MS}$
scheme throughout.

At first sight, one may argue that the first IR renormalon of the
Adler function has no significant bearing on the quantity
$r_{\tau}$, because
the singularity at $b\!=\!2$ is formally suppressed by a power of
$\alpha_s$ due to the contour integration (\ref{contint})
(see Ref.~\cite{BNP}).
We can see this, for example, if we consistently ignore
all effects beyond the one--loop in (\ref{rt2})
($\beta_j \mapsto 0$ for $j \geq 1$, $\nu \mapsto 0$).
In this approximation, the contour integration over $y$
can be carried out explicitly and it yields an oscillating function
of $b$ which has a zero at $b\!=\!2$, thus erasing the singularity there.
However, we wish to stress that this effect
implies only that the nonperturbative power term
$\sim 1/m^4_{\tau}$ contribution to
$r_{\tau}$ is
suppressed. This effect does not imply that the
behavior of the Borel transform $\widetilde D (b)$ near $b\!=\!2$
is not important for the determination of the
value of $r_{\tau}$.
In fact, if we didn't factor out the first IR renormalon singularity 
in (\ref{rt2})--(\ref{rt3}), the contributions
from the $b \sim 2$ region would be very imprecise, 
thus adversely affecting our analysis. On the other
hand, the higher IR renormalons, e.g., at $b\!=\!3$,
which are not suppressed by powers of $\alpha_s$,
contribute insignificantly to the integral (\ref{rt3a}),
as will be shown below.

\section{Predictions of \lowercase{$\alpha_s$} from the hadronic tau decay}
\label{sec:alphas}

For the evaluation of (\ref{rt3}), we will employ,
at any given  choice of $\xi^2$, 
the corresponding ${\rm N}^3 {\rm LO}$ TPS
of $R(b(w); \xi^2)$,
where we will use for $d_3^{(0)}$ the values around
$d_3^{(0)}\!=\!25.$ suggested
in Section \ref{sec:prediction}.
The double integral (\ref{rt3}) then yields,
for any given values of $\xi^2$ and $a_0\!\equiv\!a(m^2_{\tau})$,
a specific prediction for $r_{\tau}$.
We then have to adjust, at a given $\xi^2$,
the value of $a_0\!\equiv\!a(m^2_{\tau})$
in such a way that the prediction is within the
experimental limits (\ref{rtauexp}). The renormalization
scale parameter $\xi^2$ is then chosen according to the
principle of minimal sensitivity (PMS)
\begin{equation}
\frac{ \partial r_{\tau}(\xi^2) }
{\partial \xi^2} = 0 \ ,
\label{PMSeq}
\end{equation}
i.e., at the point in which the unphysical $\xi^2$--dependence
disappears locally. \footnote{It is instructive to see why our method
should fail at small and large values of $\xi^2$. At small $\xi^2$
the running coupling $a(\xi^2 Q^2)$ becomes large, and so the Borel integral
(\ref{rt3}) will receive significant contribution from the
region far beyond the first IR renormalon, in which the Borel transform
cannot be well described by the first few terms of the perturbation
theory. On the other hand, at large $\xi^2$, the coupling $a(\xi^2 Q^2)$
becomes small, and for the integral (\ref{rt3}) to be $\xi^2$-independent
the Borel transform $R(b(w);\xi^2)$
should increase rapidly as $\xi^2$ increases (In fact, it can be shown that
$R(b(w);\xi^2)$ increases approximately as $\xi^{2 b}$).
This means that the Borel transform
becomes steeper  as $\xi^2$ increases, making the
perturbation theory less efficient. It is therefore reasonable to expect
an optimal $\xi^2$ for our method, and we expect it
to be given by the PMS principle.}

\begin{figure}[ht]
 \centering\epsfig{file=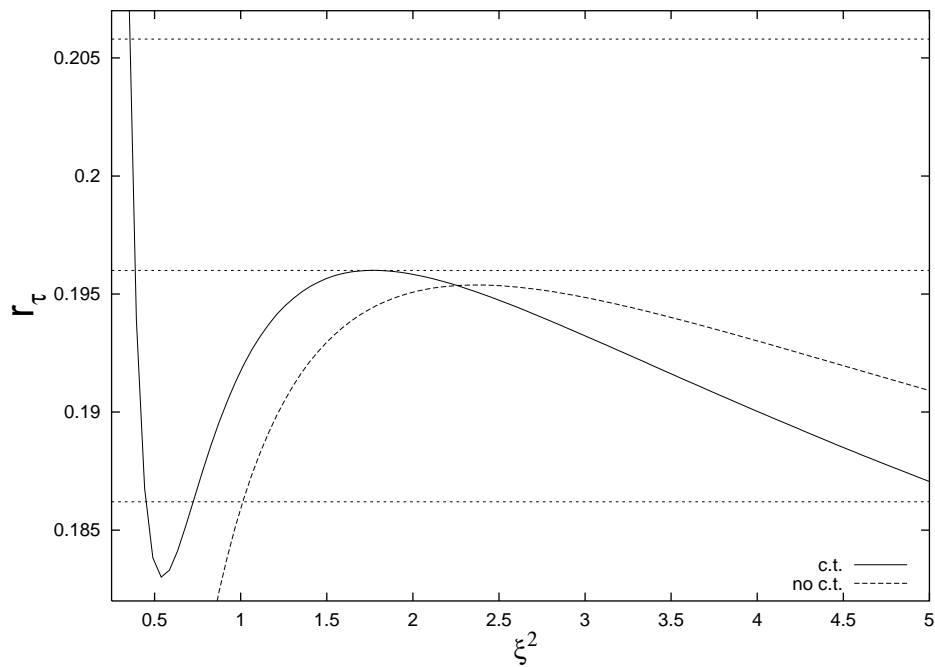}
\vspace{0.3cm}
\caption{\footnotesize
The value of the predicted ratio
$r_{\tau} \equiv 
r_{\tau}^{\rm V\!+\!A}(\triangle S\!=\!0; m_{u,d}\!=\!0)$ 
of (\ref{rt3}), 
as a function of the renormalization scale
parameter $\xi^2$, for the choice 
$\alpha_s(m^2_{\tau};{\overline {\rm MS}})\!=\!0.3265$
and $d_3^{(0)}\!=\!25.$,
when the conformal transformation (\ref{mapping}) is employed 
(full curve; $\phi\!=\!0.1$, i.e., $b_{\rm max} \approx 3.$), 
and when none is employed
(dotted curve; $b_{\rm max}\!=\!3.$). The measured
values (\ref{rtauexps}) are included as dotted horizontal lines.}
\label{fig3}
\end{figure}

There is still one minor technical detail
that we might worry about: we have only a limited
knowledge of the ${\overline {\rm MS}}$
beta function $\beta(a)$ that governs the running of the
coupling parameter $a$ -- its power expansion in $a$ is known
only up to the four--loop term 
$- \beta_0 c_3 a^5$ ($\sim\!a^5$)
\cite{RVL}. In the region with the low 
$\mu^2\!=\!\xi^2 m^2_{\tau} \exp({\rm i} y)$
($|\mu^2| \sim m^2_{\tau} \approx 3 {\rm GeV}^2$)
where the contour integration in (\ref{rt3}) is applied, 
the values of $|a|$ ($\equiv\!|\alpha_s|/\pi$) are not any
more very small ($|a| \approx 0.1$), and expansion terms with 
powers higher than $a^5$ may become significant in the
resummed value of $\beta(a)$.
To be specific, we chose the $[2/3]$
Pad\'e approximant  for the resummed
$\beta(a)$ in the RG evolution of $a$, 
above all because of the reasonable singularity structure
of this beta function ($a_{\rm singularity}\!=\!0.311$).\footnote{
This PA choice for $\beta(a)$
was motivated and used in Refs.~\cite{CK1}, where a
renormalization--scheme-- and 
scale--invariant method was developed and employed for
the resummations of NNLO TPS's of Euclidean massless QCD observables,
a generalization of the renormalization-scale--invariant Pad\'e--related
method of Refs.~\cite{CK2}.}
Later we will show how the results change when 
(${\rm N}^3 {\rm LO}$)TPS $\beta$--functions are used instead. 
Further, we chose in (\ref{rt3}) $\phi\!=\!0.1$, i.e.,
$w_{\rm max}\!=\!\exp({\rm i} \cdot 0.1)$, 
corresponding to $b_{\rm max} \approx 3.03$, i.e., well beyond the first
IR renormalon.\footnote{
When $w_{\rm max}\!=\!\exp( {\rm i} \phi)$ in (\ref{rt3}) 
is varied between 
$w(b\!=\!3)$ ($\phi \approx 0$) and 
$w(b\!=\!4)$ ($\phi \approx 0.505 \ {\rm rad}$),
the values of (\ref{rt3}) change insignificantly (relative change
is about $2.5 \cdot 10^{-6}$). \label{wmax}
}
It turns out that the $\xi^2$ values as determined by the
PMS principle (\ref{PMSeq}) of the  expression
(\ref{rt3}) are $\xi^2 \approx 1.75$--$1.80$ 
when $d_3^{(0)}\!=\!25.$
In Fig.~\ref{fig3} we show the numerical predictions
of (\ref{rt3}) as functions of the parameter
$\xi^2$, for the choice 
$\alpha_s^{(0)}\!\equiv\!\alpha_s(m^2_{\tau}) 
= 0.3265$ (and $d_3^{(0)}\!=\!25.$). 
The central experimental value (\ref{rtauexp})
$r_{\tau}\!=\!0.1960$
is then achieved at the PMS (\ref{PMSeq}) value $\xi^2 \approx 1.77$.
We see that the
unphysical $\xi^2$--dependence is really quite weak
in a large interval $1. < \xi^2 < 5.$, indicating that
the method is reliable. In the Figure, we include for 
comparison the analogous predictions for the case (\ref{rt2}),
i.e., when
no conformal transformation $b \mapsto b(w)$ is
carried out in (\ref{rt2}) [we used $\varepsilon\!=\!0.005$ 
and $b_{\rm max}\!=\!3$ in (\ref{rt2})]. 
The latter method has a somewhat different $\xi^2$--dependence
and a slightly different value at the PMS point. As argued after
Eq.~(\ref{rt2}), the predictions of the curve(s)
involving the conformal transformation are expected to be
more reliable.

The predictions for 
$\alpha_s(m^2_{\tau})$,
obtained by matching the results of the described
resummation (\ref{rt3}) with the experimental results
(\ref{rtauexp}), for the choice $d_3^{(0)}\!=\!25.$, are
\begin{equation}
\alpha_s(m^2_{\tau}) = 
0.3265 \pm 0.0062_{\rm exp.} \pm 0.0062_{\rm EW} \pm 0.0053_{\rm CKM}
\quad (d_3^{(0)} = 25.) \ .
\label{res1}
\end{equation}
The perturbative QCD part
of information incorporated in the prediction (\ref{res1})
was the ${\rm N}^3 {\rm LO}$ TPS for the $N_f\!=\!3$
Adler function $D(Q^2)$, with the ${\rm N}^3 {\rm LO}$ coefficient
$d_3(\xi^2=1)\!\equiv\!d_3^{(0)}$
set equal to $d_3^{(0)}=25.$, as obtained by the arguments
of Sec.~\ref{sec:prediction}. Of course, the exact value of 
$d_3^{(0)}$ is not yet known. 
The authors of Ref.~\cite{KS}, using the
effective charge (ECH) \cite{ECH,KKP,Gupta}
and the TPS principle of minimal sensitivity (PMS) \cite{PMS,PMS2}
methods, predicted $d_3^{(0)}\!=\!27.5$.
When considering a one--parameter subgroup $Q^2 \mapsto e^{\gamma} Q^2$
of the renormalization group, 
which of course leaves the 
coefficients of the (ECH) $\beta$--function
$d [D(Q^2)]/ d [\ln Q^2] = - \beta_0 d^2 (1 + \rho_1 d + \rho_2 d^2 + ...)$
invariant, the authors of Ref.~\cite{GKP} obtained
an estimate $d_3^{(0)}\!=\!30.9$ using a variant of the
PMS, and the authors of Ref.~\cite{CKT} obtained
$d_3^{(0)}\!=\!28.7$ using a so--called G-scheme.
Further, when employing the simple $[2/1]$ PA estimate
for the NNLO TPS $D(Q^2)=a_0 (1 + d_1^{(0)} a_0 + d_2^{(0)} a_0^2)$,
at $\mu^2\!=\!Q^2$ ($=\!m_{\tau}^2$), the prediction is
$d_3^{(0) {\rm pr.}} = d_2^{(0) 2}/d_1^{(0)} = 24.75$.
If using the simple $[3/1]$ PA estimate for the ECH TPS 
$\beta$--function $- \beta_0 d^2 (1 + \rho_1 d + \rho_2 d^2)$,
the prediction is $\rho_3^{\rm pr.} = \rho_2^2/\rho_1 = 5.39$
and thus\footnote{
Note that 
$\rho_1\!=\!- d_1^{(0)}\!+\!\beta_0 \ln m_{\tau}^2/{\widetilde \Lambda}^2
\!=\!5.094$ is obtained here by using
the (unsubtracted) Stevenson equation \cite{PMS}, with 
$\alpha_s(m_{\tau}^2)\!=\!0.33$
and with the $[2/3]$ PA for the
${\overline {\rm MS}}$ $\beta$--function;
$\rho_2\!=\!d_2^{(0)}\!-\!d_1^{(0) 2}\!-\!c_1 d_1^{(0)}\!+\! 
c_2^{\overline {\rm MS}}\!=\!5.238$. Further,
$d_3^{(0) {\rm pr.}}\!=\!\rho_3^{\rm pr.}\!+\!d_1^{(0)} [
2 d_2^{(0)}\!-\!d_1^{(0) 2}\!+\!c_1 d_1^{(0)}/2\!+\!\rho_2] -
c_3^{\overline {\rm MS}}/2$.}
$d_3^{(0) {\rm pr.}} = 22.4$.
Keeping all these estimates for $d_3^{(0)}$ in mind,
as well as the estimate (\ref{d3pr}) of our approach, 
it appears reasonable and safe
to allow for the following variation 
of the values of $d_3^{(0)}$ around the value $25.$ 
from Sec.~IV:
\begin{equation}
d_3^{(0) {\rm est.}} = 25. \pm 10. 
\label{d3o0var}
\end{equation}
The $\pm 10.$ variation in $d_3^{(0)}$ results in the 
variation of $\mp 0.0039$ for 
$\alpha_s(m^2_{\tau})$, respectively.
The $\xi^2$ values as determined by the PMS principle
(\ref{PMSeq}) vary as well: $\xi^2_{\rm PMS}
\approx 2.10, 1.75, 1.35$ for $d_3^{(0)}\!=\!15., 25., 35.$,
respectively.

In order to obtain an estimate of the various
uncertainties in $\alpha_s(m^2_{\tau})$
due to the use of the method itself,
we proceed the following way. 

One of the major uncertainties is connected with
our truncation of $R(b(w),\xi^2)$ 
to the ${\rm N}^3 {\rm LO}$ TPS.
One way of estimating these uncertainties
would be to repeat the analysis with using
${\rm N}^4 {\rm LO}$ TPS for $R(b(w),\xi^2)$ in (\ref{rt3}).
For this, we need also the value of the coefficient
$d_4^{(0)}$ in $D(Q^2)$.
We note that the coefficients $d_j^{(0)}$ in
$D(Q^2)$ follow roughly the geometric series pattern,
with $d_2^{(0)}/d_1^{(0)} \approx d_3^{(0)}/d_2^{(0)} \approx 4$.
Therefore, we may estimate $d_4^{(0)} \approx 4 d_3^{(0)}$.
Using these values of $d_4^{(0)}$, with the $d_3^{(0)}$
values (\ref{d3o0var}), our method gives
predictions for $\alpha_s(m^2_{\tau})$
which differ from the original (${\rm N}^3 {\rm LO}$ TPS) 
method by $0.0012, 0.0007, 0.0003$ when
$d_3^{(0)}\!=\!15.,25.,35.$, respectively.
The PMS-determined $\xi^2$ are in the ${\rm N}^4 {\rm LO}$ TPS case
$\xi^2_{\rm PMS} \approx 3.2, 2.7, 2.15$, respectively.
However, if we fix $\xi^2$ to the PMS-determined values
of the original (${\rm N}^3 {\rm LO}$ TPS) method
($2.10, 1.75, 1.35$, respectively), then the
differences in the predictions for $\alpha_s(m^2_{\tau})$
are $0.0035, 0.0025, 0.0020$, respectively.
Choosing the largest difference here, this would
suggest that the truncation uncertainty in our
prediction of $\alpha_s(m^2_{\tau})$ is about 0.0035.

We may obtain another estimate of the truncation
error in the following way.
We use for $R(b(w))$ in (\ref{rt3}), instead
of the ${\rm N}^3 {\rm LO}$ TPS of the type
(\ref{tbxi2=1})--(\ref{tbxi2=2}), the corresponding Pad\'e
approximant (PA) $[2/1](w)$.
We expect the most reasonable pole of this PA
to be $w_{\rm pole} \approx 1$, corresponding to
$b \approx 3$ (i.e., the second IR renormalon pole).
We vary $d_3^{(0)}$, at two fixed values of $\xi^2$--parameter:
$\xi^2\!=\!1.75$, and $1.95$ (i.e., $\approx\!\xi^2_{\rm PMS}$
for $d_3^{(0)}\!=\!25., 20.$, respectively)
in such a way that $w_{\rm pole}$ varies between
$w_{\rm pole}\!=\!1$ and $w_{\rm pole}\!=\!0.64$.
The latter value corresponds to the location of the $b$--pole 
half--way between the first and the second renormalon
[$b(w\!=\!0.64) \approx 2.5$]. 
The variations of $d_3^{(0)}$ needed for this
are $d_3^{(0)}\!=\!23.5$--$27.0$, and $21.0$--$25.5$,
respectively. The variation of the predictions of
$\alpha_s(m^2_{\tau})$ for such variation of $d_3^{(0)}$,
with the use of ${\rm N}^3 {\rm LO}$ TPS and $[2/1]$ PA
for $R(b(w),\xi^2)$, is then
$\delta \alpha_s(m^2_{\tau}) = 0.0042, 0.0048$,
for the two aforementioned choices of $\xi^2$, respectively.
This variation (e.g., the larger one: 0.0048) can be
regarded as an estimate of the truncation error
of our method, especially since the PA $[2/1]_R(w)$
represents a specific realization of the resummation
of $R(b(w),\xi^2)$. Since this estimate is larger than
the previous one (0.0035), we will use it:
$\delta \alpha_s(m^2_{\tau})_{\rm tr.} = 0.0048$.

There is also an uncertainty in the predictions
of our method due to possible ambiguities in
the choice of the renormalization scale parameter $\xi^2$.
Our choice was to fix $\xi^2$
by the local PMS principle (\ref{PMSeq}).
Somewhat similarly as we estimated the truncation error,
we may now vary $\xi^2$ instead and keep $d_3^{(0)}$
fixed ($=\!25.$). If we vary $\xi^2$ 
from  $\xi^2 \approx 1.55$
to $\xi^2 \approx 2.0$,
the aforementioned PA $[2/1]_R(w)$
changes its pole from $w_{\rm pole}\!=\!1$ to
$w_{\rm pole}\!=\!0.64$.
The resulting variation in the predicted (central) values of 
$\alpha_s(m^2_{\tau})$, with the use of 
${\rm N}^3 {\rm LO}$ TPS and $[2/1]$ PA
for $R(b(w),\xi^2)$, is then about $0.0033$. 
Alternatively, the change $-0.0033$
in $\alpha_s(m^2_{\tau})$ would correspond to the
variation of the renormalization scale parameter
$\xi^2 \approx 1.50$--$4.10$ around its PMS (\ref{PMSeq}) value
$\xi^2 \approx 1.77$ when the ${\rm N}^3 {\rm LO}$ TPS
approach of (\ref{rt3}) is applied.
Very similar results are obtained if $d_3^{(0)}\!=\!20.$
is used instead.
We will take for the uncertainty due to
the $\xi^2$--ambiguity the value
$\delta \alpha_s(m^2_{\tau})_{\delta \xi^2} = 0.0033$. 

Further, the predictions change when the
renormalization scheme parameters $c_2$ and $c_3$ change.
The leading scheme parameter is $c_2$. We have
$c_2^{\overline {\rm MS}}\!=\!4.471$ and
$c_3^{\overline {\rm MS}}\!=\!20.99$ ($N_f\!=\!3$).
For comparison, for the ${\rm N}^3 {\rm LO}$ TPS Adler function
(with $d_3^{(0)}\!=\!25.$) 
in the TPS PMS scheme \cite{PMS} we have
$c_2^{\rm PMS}\!=\!6.584$, $c_3^{\rm PMS}\!=\!36.80$
(and $\xi^2 \approx 0.55$),
and in the ECH scheme \cite{ECH,KKP,Gupta}
$c_2^{\rm ECH}\!=\!5.238$, $c_3^{\rm ECH}\!=\!16.06$
(and $\xi^2 \approx 0.48$).
This would indicate that it is reasonable to
allow for the variation of the leading scheme
parameter $c_2$ from its ${\overline {\rm MS}}$
value by about $50 \%$, i.e., $c_2 = 4.471 (1. \pm 0.5)$,
while adjusting the renormalization scale parameter
$\xi^2$ according to the PMS condition (\ref{PMSeq}).
The central prediction in (\ref{res1}) then varies
by about $\pm 0.0019$.
On the other hand, changing the NNLO scheme parameter 
$c_3$ by $50 \%$ around its  ${\overline {\rm MS}}$ value
changes the central prediction for $\alpha_s(m^2_{\tau})$
by about $\pm 0.0006$. Adding in quadrature, the
uncertainty due to the change of the scheme
parameters\footnote{
The problem of the renormalization scale and scheme dependence
in the determination of $\alpha_s(m^2_{\tau},{\overline {\rm MS}})$
from the $y$--contour representation (\ref{contint}) of
$r_{\tau}$
was discussed by R\c aczka \cite{Raczka}. Using the
NNLO TPS for $D(Q^2)$, he showed that a change from
the ${\overline {\rm MS}}$ scheme (with $\xi^2\!=\!1$)
to the TPS PMS scheme and scale results in the change
of $\alpha_s(m^2_{\tau},{\overline {\rm MS}})$ by $0.01$,
which is significant in the view of the new precise
experimental data.}
is about $\pm 0.0020$.

Hence, our result is:
\begin{eqnarray}
\alpha_s(m^2_{\tau})&= &  
0.3265 \pm 0.0062_{\rm exp.} \pm 0.0062_{\rm EW} \pm 0.0053_{\rm CKM} \pm
\nonumber\\
&&0.0039_{\delta d_3} \pm 0.0048_{\rm tr.} 
\pm 0.0033_{\rm \delta \xi^2} \pm 0.0020_{\delta c_j} \ ,
\label{res2}
\\
&=& 0.3265 \pm 0.0062_{\rm exp.} \pm 0.0082_{\rm EW + CKM} 
\pm 0.0073_{\rm meth.}\ .
\label{res3}
\end{eqnarray}
In the last line, we added the corresponding uncertainties 
in quadrature; the method uncertainty contains the
uncertainties due to the variation of $d_3^{(0)}$,
truncation error, and renormalization scale and scheme
ambiguities. If we use for the ${\overline {\rm MS}}$ $\beta$--function
the ${\rm N}^3 {\rm LO}$ TPS, instead of $[2/3]_{\beta}$,
the predictions in (\ref{res2})--(\ref{res3}) decrease
by $0.0006$--$0.0007$, indicating that those nonperturbative
effects which originate in the behavior of the 
$\beta$--function are not strong in the applied
resummation method. This has to do with relatively large 
values of the PMS--fixed (\ref{PMSeq}) renormalization scale
parameters $\xi^2\!\equiv\!\mu^2/m^2_{\tau}$
($\approx 1.77$ when $d_3^{(0)}\!=\!25.$).
The uncertainty due to the variation of
$b_{\rm max}$ in (\ref{rt3a}) 
($3.\!<\!b_{\rm max}\!<\!4.$, i.e., 
$0\!<\!{\phi}\!<\!0.505 \ {\rm rad}$)
turns out to be insignificant, as mentioned before. 
For example, if changing
from $\phi=0.1$ (corresponding to $b_{\rm max} \approx 3.03$) 
to $\phi = 0.505$ ($b_{\rm max} \approx 4.$) 
the central value of (\ref{res3}) increases by
less than $10^{-6}$.

One may argue that the method uncertainty as given above
is too nonconservative, i.e., too small.
Therefore, we carried out an additional cross--check.
We performed the resummation for $r_{\tau}$
by the double integration of the type
(\ref{rt3}), but this time taking into
account explicitly the first UV renormalon
(at $b\!=\!-1$) in the ansatz for the Borel transform:
$\widetilde D(b) = \overline{R}(b) (1 + b)^{- \gamma_1}
(1-b/2)^{-\left(1+\nu\right)}$, with $\gamma_1=2.589$ \cite{BBK,caprini}.
We performed again the conformal mapping $b=b(w)$ (\ref{mapping}),
expansion of $\overline R (b(w))$ in powers of $w$
up to and including the ${\rm N}^3 {\rm LO}$ ($\sim\!w^3$),
and subsequently performed the double integration
analogous to (\ref{rt3}), with $\phi=0.1$. The scale parameter $\xi^2$
was again fixed by the PMS principle (\ref{PMSeq}),
resulting, for the choices $d_3^{(0)}=15., 25.,35.$ (and 
$\alpha_s(m^2_{\tau}) \approx 0.32$--$0.34$)
in considerably lower values $\xi^2 \approx 0.91, 0.88, 0.85$,
respectively. We used for the ${\overline {\rm MS}}$
$\beta$--functions again the PA $[2/3]$.
The predicted values of the QCD coupling parameter
turned out to be very close to those (\ref{res2})--(\ref{res3})
of the method (\ref{rt3}):
$\alpha_s(m^2_{\tau}) = 
0.3257 \pm 0.0062_{\rm exp.} \pm 0.0062_{\rm EW} \pm 0.0054_{\rm CKM}
\pm 0.0014_{\delta d_3}$.
For example, the prediction for
$\alpha_s(m^2_{\tau})$
corresponding to the
central experimental value of (\ref{rtauexp})
with this method differed from the prediction
of the method (\ref{rt3}) by $-0.0027$, $-0.0008$, $+0.0018$,
when $d_3^{(0)}=15., 25., 35.$,
respectively. This would indicate again that the
resummation method uncertainty 
does not surpass $0.0073$, i.e., in accordance with
the method uncertainty estimate in (\ref{res3}).

If we apply the resummation (\ref{rt3a})
{\em without\/} the conformal transformation 
[using $b_{\rm max} \approx 3$,  $\varepsilon= 0.005$ in (\ref{rt2})],
for $d_3^{(0)}\!=\!15., 25.,35.$,
the PMS--fixed (\ref{PMSeq}) renormalization scale parameters are
$\xi^2_{\rm PMS} \approx 3.00, 2.35, 1.25$, respectively,\footnote{
For $d_3^{(0)}\!=\!35.$, no strict stationarity is achieved,
but at $\xi^2 \approx 1.25$ the slope (\ref{PMSeq}) is almost
zero: $\partial r_{\tau}(\xi^2)/ \partial \xi^2 \approx 
- 2.3 \cdot 10^{-4}$.}
and the prediction is
$\alpha_s(m^2_{\tau})\!=\!
0.3271 \pm 0.0062_{\rm exp.} \pm 0.0062_{\rm EW} \pm 0.0053_{\rm CKM}
\pm 0.0060_{\delta d_3}$,
which\footnote{
The variation $\pm 0.0060_{\delta d_3}$ for $d_3^{(0)}\!=\!25. \pm 10.$
is in fact $+ 0.0043$ for $d_3^{(0)}\!=\!15.$, and $- 0.0060$ for
$d_3^{(0)}\!=\!35.$} 
is only slightly different from the one
with the conformal transformation (\ref{res1})--(\ref{res2}).
Although not using the conformal transformation 
is not so well motivated
[see also the discussion after (\ref{rt2})], this 
result may represent yet another
justification for the small estimate of the
method uncertainty in (\ref{res2})--(\ref{res3}).

The result (\ref{res3}) was then evolved from
the scale $\mu=m_{\tau} \approx 1.777$ GeV
to the canonical scale $M_{\text{z}}\!=\!91.19$ GeV, 
by using the RG equation with the 
$[2/3]_{\beta {\overline {\rm MS}}}$  Pad\'e approximant 
(based on the four--loop $\beta_{\overline {\rm MS}}$ \cite{RVL})
and the three-loop matching conditions \cite{CKS}
for the flavor thresholds. We used the matching at
$\mu(N_f)\!=\!\kappa m_q(N_f)$ with the choice $\kappa\!=\!2$,
where $\mu(N_f)$ is the scale
above which $N_f$ flavors are assumed active, and
$m_q(N_f)$ means the running quark mass $m_q(m_q)$
of the $N_f$'th flavor [we assumed $m_c(m_c)\!=\!1.25$ GeV
and $m_b(m_b)\!=\!4.25$ GeV]. This leads to our final result
\begin{eqnarray}
\alpha_s(M^2_{\text{z}}) &=& 0.1193 \pm 0.0007_{\rm exp.}
\pm 0.0010_{\rm EW+CKM}
\pm 0.0009_{\rm meth.} \pm 0.0003_{\rm evol.} \ ,
\label{res4}
\\
& = & 0.1193 \pm 0.0015 \ .
\label{res5}
\end{eqnarray}
In (\ref{res5}), we added the uncertainties of (\ref{res4})
in quadrature. 
In (\ref{res4}), we included the uncertainty 
$\pm 0.0003_{\rm evol.}$ due to the RG evolution
from $m_{\tau}$ to $M_{\text{z}}$. This uncertainty
estimate is obtained in the following way.
Keeping $\kappa\!=\!2$, if we vary the mass 
$m_c(m_c)\!=\!1.25 \pm 0.10$ GeV, the resulting uncertainty
is $\pm 0.0002$; if we vary the mass
$m_b(m_b)\!=\!4.25 \pm 0.15$ GeV, the
uncertainty is $\pm 0.0001$.
If we vary the 
flavor threshold parameter $\kappa$ around its
central value $\kappa\!=\!2$ from $1.5$ to $3.$,
the uncertainty is $\pm 0.0001$.
Furthermore, if we use for the ($m_{\tau}\!\to\!M_{\text{z}}$)
RG evolution, instead of the PA $[2/3]_{\beta {\overline {\rm MS}}}$, 
the corresponding four-loop TPS $\beta_{\overline {\rm MS}}$ function, 
the resulting
$\alpha_s(M_{\text{z}}^2)$ changes by $0.0001$.
Adding all these uncertainties in quadrature
gives us approximately the uncertainty $\pm 0.0003$
given in (\ref{res4}).

If we repeat the entire calculation of $\alpha_s(m^2_{\tau})$
and $\alpha_s(M^2_{\rm z})$ by using throughout the four--loop
TPS ${\overline {\rm MS}}$ $\beta$--function instead of
the PA $[2/3]_{\beta {\overline {\rm MS}}}$, 
the predictions for $\alpha_s(M^2_{\rm z})$ remain the same
as in (\ref{res4})--(\ref{res5}), up to the displayed digits.
The reason for this is that the aforementioned
change of $\beta$--functions predicts the
values of $\alpha_s(m^2_{\tau})$ by about
$0.0006$--$0.0007$ lower than those of
(\ref{res2})--(\ref{res3}), but then the
RG evolution to $\mu\!=\!M_{\rm z}$ pushes
the results up, thus approximately neutralizing
this effect. 

Due to the high precision experimental data 
(\ref{Rtauexp})--(\ref{rtauexp}) on the inclusive
hadronic decay of $\tau$, the experimental
uncertainty in the extracted strong
coupling constant is low. By incorporating
a wealth of known theoretical information
(perturbative as well as renormalon)
on the related Adler function $D(Q^2)$, we were able
to extract the strong coupling constant
with the method uncertainty not significantly surpassing
the experimental uncertainty. 
Further, the analysis by the ALEPH Collaboration 
\cite{ALEPH1,Davier} 
showed that those power (nonperturbative) contributions 
in the observable 
$R_{\tau}^{\rm V\!+\!A}(\triangle S\!=\!0)$
which do not originate from the nonzero quark masses
are consistent with zero (see also the next Section),
and these OPE--type contributions were not included
in our resummation either.

The experimental situation with
other low energy QCD observables is not so favorable,
and the experimental uncertainties of the extracted 
strong coupling constants appear to dominate over the
theoretical uncertainties. This is reflected in the present 
world average (over various measured QCD observables)
${\alpha}_s^{{\overline {\rm MS}}}(M_{\text{z}}^2) =
0.1173 \pm 0.0020$ by Ref.~\cite{Hinchliffe} and
$0.1184 \pm 0.0031$ by Ref.~\cite{Bethke},
where the extracted (combined experimental and theoretical)
uncertainties are significantly
higher than those in (\ref{res4}).

In this context, we mention that the question
of the violation of the quark(gluon)--hadron duality 
for correlation functions has been raised and
investigated by the authors of Refs.~\cite{duality}. 
They argued that the corrections to the correlation functions
due to the duality violation could be significant
(up to a few percent). However, no quantitative analyses
are available at present. This violation could possibly affect
many QCD (quasi)observables, including the Adler function $D(Q^2)$
and $R_{\tau}$.

Further, the authors of Refs.~\cite{CNZ,HRS} analyzed the
possibility that the Operator Product Expansion (\ref{OPE})
contains an additional $1/Q^2$--term (other than the
$d\!=\!2$ quark mass terms), whose origin
would be an effective tachyonic gluon mass reflecting
short--distance nonperturbative QCD effects.
The authors
of Ref.~\cite{CNZ} suggested that such terms would decrease
the value of $\alpha_s(m^2_{\tau})$ extracted from hadronic
$\tau$ decays by about $10 \%$. However, 
the authors of Ref.~\cite{DoSch} showed that
the coefficient of the $1/Q^2$--term is consistent with zero.
They did this by fitting a dimension--two finite energy sum
rule to the new ALEPH data on the vector and axial--vector
spectral functions extracted from measured $\tau$ decays.
The type of the sum rule used by the authors of Ref.~\cite{DoSch}
in ruling out the aforementioned $1/Q^2$--term are well
satisfied at the continuum threshold scales
$s_0 \approx 2$--$3 \ {\rm GeV}^2$ relevant for $r_{\tau}$, 
as has been shown independently by two groups
\cite{gr1,gr2}.

\section{Comparison with other analyses}
\label{sec:comparison}

We may compare our result (\ref{res3})
with that of an independent analysis
of the hadronic $\tau$ decays
by the ALEPH Collaboration \cite{ALEPH1,Davier},
who used for $R_{\tau}^{\rm V+A}(\triangle S\!=\!0)$, 
instead of the values (\ref{Rtauexp}),
the different values available in 1998
($3.492 \pm 0.016$)
\begin{eqnarray}
\alpha_s(m^2_{\tau}) &=&
0.334 \pm 0.007_{\rm exp.} \pm 0.021_{\rm th.} \quad
({\rm ALEPH}) \ , \quad \Rightarrow
\label{resALmt}
\\
\alpha_s(M^2_{\text{z}}) &=&
0.1203 \pm 0.0008_{\rm exp.} \pm 0.0025_{\rm th.} 
\pm 0.0003_{\rm evol.} \quad
({\rm ALEPH}) \ .
\label{resALMZ}
\end{eqnarray}
They used slightly smaller uncertainties for $\delta_{\rm EW}$
($\pm 0.0040$), and drastically smaller uncertainties
in the CKM element: $|V_{ud}|\!=\!0.9752 \pm 0.0007$.
They used different methods which involved, in addition,
the analysis of moments of (their own measured)
spectral functions ${\rm Im} \Pi^{(J)}_{{\bar u}d, V/A}(s)$
($s \leq m^2_{\tau}$) as proposed by \cite{LDP2}.
The ALEPH's ${\rm V\!+\!A}$ analysis of the mentioned moments
showed that those nonperturbative (OPE) contributions 
which do not originate from the nonzero quark masses
were consistent with zero:
$\delta r_{\tau}({\rm NP}, m_{u,d}\!=\!0) = 0.000 \pm 0.004$;
and their quark mass nonperturbative
contributions basically agree with ours (\ref{drtmud2})
(compare Table 8, fourth column, of Ref.~\cite{ALEPH1}).
Further, the central value of (\ref{resALmt}) was obtained
by taking the arithmetic average of the predictions of
two methods: 1.) the $y$--contour integration
approach (\ref{contint}), with just the ${\rm N}^3 {\rm LO}$ TPS
for the Adler $D(Q^2)$ function, with $d_3^{(0)}\!=\!50. \pm 50.$,
in ${\overline {\rm MS}}$, i.e., the approach of \cite{rtau6};
2.) the simple ${\rm N}^3 {\rm LO}$ TPS of the power expansion
of $r_{\tau}$.
The large theoretical uncertainty in (\ref{resALmt})
originates primarily from the difference of the 
predictions of the two aforementioned methods,
from the ambiguities of the choice of $d_3^{(0)}$, 
renormalization scheme and scale, and the electroweak
parameter $\delta_{\rm EW}$.

The $y$--contour integration
approach where we just use the ${\rm N}^3 {\rm LO}$ TPS
for the Adler function $D(Q^2)$ (with $d_3^{(0)}\!=\!25.$)
in the contour integral (\ref{contint}),
i.e., the approach of \cite{rtau6}, in ${\overline {\rm MS}}$
scheme with ${\rm N}^3 {\rm LO}$ TPS $\beta_{\overline {\rm MS}}$, 
achieves the minimal sensitivity
condition (\ref{PMSeq}) at $\xi^2 \approx 0.4$
and reproduces there the central experimental value
of (\ref{rtauexp}) at 
$\alpha_s(m^2_{\tau}) = 0.3399$ ($0.3416$ if using 
$[2/3]_{\beta {\overline {\rm MS}}}$), 
corresponding to
$\alpha_s(M^2_{\text{z}}) = 0.1209$ ($0.1210$),
significantly higher\footnote{
This contour approach, in ${\overline {\rm MS}}$ scheme, 
was also applied in Ref.~\cite{GKPMPLA}. On the other hand,
if we apply in this approach the 
${\rm N}^3 {\rm LO}$ TPS Adler function
in the ECH renormalization scale and scheme 
($\xi^2_{\rm ECH}\!=\!0.482483$; $c_2^{\rm ECH}\!=\!5.23783$;
$c_3^{\rm ECH}\!=\!16.0613\pm 10.$ for $d_3^{(0)}\!=\!25.\pm 10.$),
and using the  ${\rm N}^3 {\rm LO}$ TPS (or: $[2/3]$ Pad\'e approximants) 
for the $\beta$--functions,
we obtain from (\ref{rtauexp}) similar results: 
$\alpha_s(m^2_{\tau},{\overline {\rm MS}}) = 0.3400 \pm 0.0080_{\rm exp.}
\pm 0.0080_{\rm EW} \pm 0.0069_{\rm CKM} \pm 0.0034_{\delta d_3}$
($0.3413 \pm 0.0083_{\rm exp.}
\pm 0.0083_{\rm EW} \pm 0.0071_{\rm CKM} \pm 0.0034_{\delta d_3}$), 
corresponding to
$\alpha_s(M^2_{\rm z},{\overline {\rm MS}}) = 0.1210
\pm 0.0009_{\rm exp.} \pm 0.0012_{\rm EW+CKM}
\pm 0.0004_{\delta d_3} \pm 0.0003_{\rm evol.}$.}
than our central value (\ref{res4}). 
However, if taking instead the simple TPS
$r_{\tau} = 
a [1 + (d_1^{(0)}\!+\!3.563) a + (d_2^{(0)}\!+\!19.99) a^2
+ (d_3^{(0)}\!+\!78.) a^3]$ (see \cite{rtau6}), at renormalization
scale $\mu\!=\!m_{\tau}$, the predictions change significantly: 
$\alpha_s(m^2_{\tau}) = 
0.3211 \pm 0.0056_{\rm exp.} \pm 0.0056_{\rm EW} \pm 0.0048_{\rm CKM}
\pm 0.0011_{\delta d_3}$,
(for $d_3^{(0)} = 25. \pm 10$.), corresponding to
$\alpha_s(M^2_{\text{z}}) = 0.1188 \pm 0.0007_{\rm exp.}
\pm 0.0010_{\rm EW+CKM} \pm 0.0002_{\delta d_3}$. 
These values are lower than our values (\ref{res4}).
The arithmetic average
of the central values of these two methods
$\alpha_s(m^2_{\tau})_{\rm arithm.} =
(0.3399\!+\!0.3211)/2 = 0.3305$
is close to the central value of the ALEPH analysis (\ref{resALmt}),
the difference being that the ALEPH used the higher values
for the hadronic $\tau$ decay ratio available in 1998. 
This value is close
to the upper bounds of our prediction (\ref{res3}).

\begin{figure}[ht]
 \centering\epsfig{file=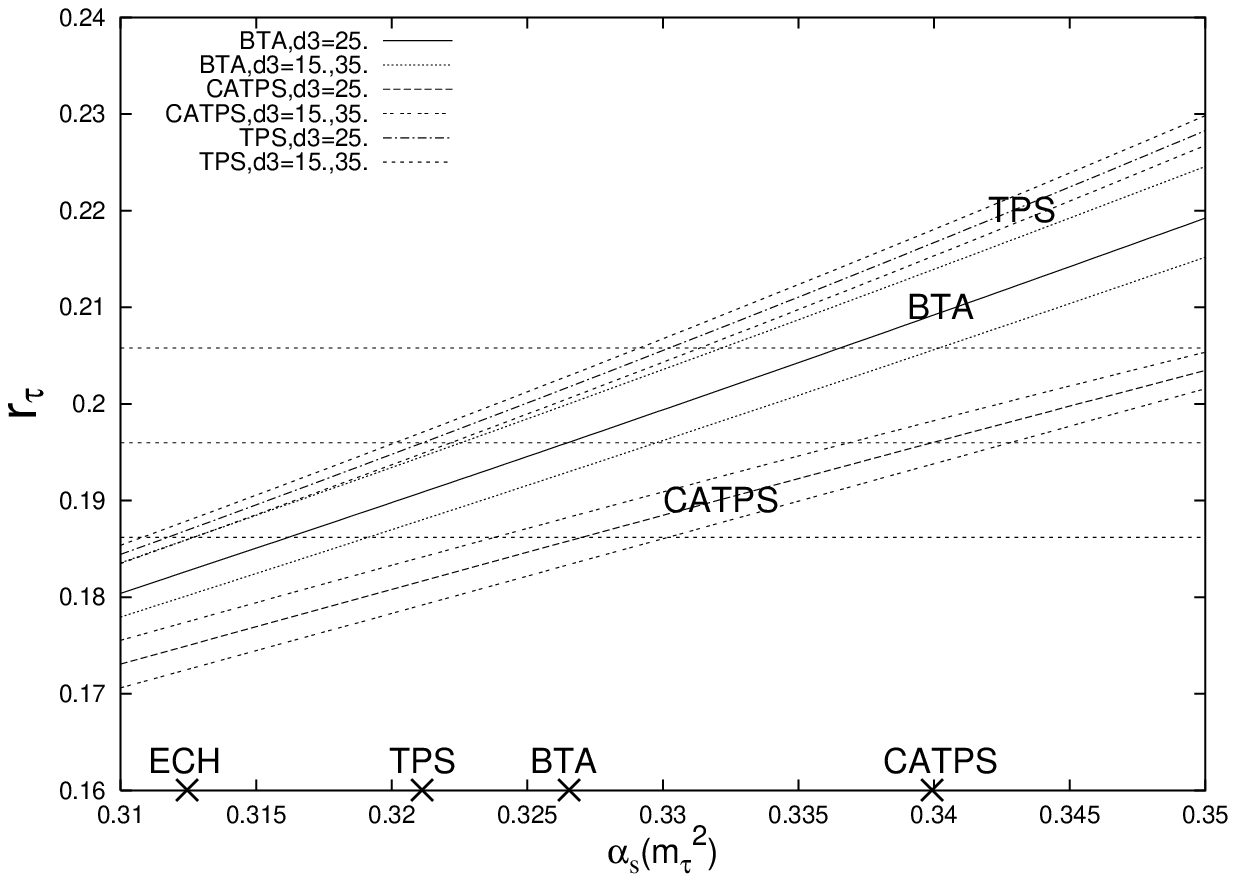}
\vspace{0.3cm}
\caption{\footnotesize
The values of the predicted ratio
$r_{\tau} \equiv 
r_{\tau}^{\rm V\!+\!A}(\triangle S\!=\!0; m_{u,d}\!=\!0)$
as functions of ${\alpha}_s(m^2_{\tau},{\overline {\rm MS}})$, from
various methods: our Borel transform approach of (\ref{rt3}) 
[BTA, with $\phi=0.1$ and PMS condition (\ref{PMSeq})]; 
the contour approach method using the ${\rm N}^3 {\rm LO}$ TPS for
the Adler function (CATPS, with the ``PMS'' $\xi^2=0.40$);
the fixed ${\rm N}^3 {\rm LO}$ TPS evaluation 
of $r_{\tau}$ (TPS, at $\mu^2\!=\!m^2_{\tau}$).
The uncertainties due to $d_3^{(0)}\!=\!25.\pm 10.$ are included. 
The measured values (\ref{rtauexps}) are included as 
dotted horizontal lines. On the $x$--axis, we denoted the values
of ${\alpha}_s(m^2_{\tau},{\overline {\rm MS}})$ of these three
methods (with $d_3^{(0)}\!=\!25.$)
for which the central measured value 
$r_{\tau}\!=\!0.1960$ is obtained.
In addition, we included the analogous prediction of the 
(NNLO) ECH method when applied to the fixed NNLO TPS of $r_{\tau}$.}
\label{fig4}
\end{figure}

In Fig.~\ref{fig4} we present the predictions of three methods
for $r_{\tau}$ as functions
of $\alpha_s(m^2_{\tau})$: 1.) our method
(\ref{rt3}) (Borel transform approach -- BTA, in
$[2/3]_{\beta}$ ${\overline {\rm MS}}$ scheme);
2.) the aforementioned method of the $y$--contour approach for
the ${\rm N}^3 {\rm LO}$ TPS Adler function (CATPS) (with
``PMS'' $\xi^2\!=\!0.40$, in ${\rm N}^3 {\rm LO TPS}_{\beta}$
${\overline {\rm MS}}$ scheme);
3.) the ${\rm N}^3 {\rm LO}$ TPS of 
$r_{\tau}$ (with $\mu^2\!=\!m^2_{\tau}$).
The values (\ref{d3o0var}) were taken for 
the ${\rm N}^3 {\rm LO}$ coefficient 
$d_3^{(0)}\!\equiv\!d_3(\mu^2\!=\!Q^2,{\overline {\rm MS}})$.
On the $x$--axis, we denoted the corresponding values
of $\alpha_s(m^2_{\tau})$ when the predicted values of the
three methods, with $d_3^{(0)}\!=\!25.$, reach the central
experimental value (\ref{rtauexp})
$r_{\tau}\!=\!0.1960$.

In addition, we included on $x$--axis of the Figure
the central value of $\alpha_s(m^2_{\tau})$ predicted 
by the (NNLO) ECH approach \cite{ECH,KKP,Gupta}
applied directly to the
NNLO TPS of $r_{\tau}$ (with using the
NNLO TPS $\beta$--functions of the ECH and ${\overline {\rm MS}}$
schemes: $c_3^{\rm ECH}, c_3^{\overline {\rm MS}} \mapsto 0$),
i.e., the approach applied in Ref.~\cite{KoKP}.
The authors of Ref.~\cite{KoKP} used the
input values $r_{\tau} = 0.2030 \pm 0.0070_{\rm exp.}$,
which yield 
$\alpha_s(m^2_{\tau}) = 0.3184 \pm 0.0060_{\rm exp.}$ and
$\alpha_s(M^2_{\rm z}) = 0.1184 \pm 0.0007_{\rm exp.}$.\footnote{
The evolution uncertainty $\pm 0.0006_{\rm evol.}$ given
in Ref.~\cite{KoKP} is larger than ours in (\ref{res4}), 
possibly because they used a lower threshold parameter
$\kappa\!=\!1$, while we used $\kappa_{\rm central}\!=\!2.$ and
$\kappa\!=\!1.5$--$3$.}
Using our updated input values (\ref{rtauexp}) for $r_{\tau}$,
this method predicts 
$\alpha_s(m^2_{\tau}) = 0.3124 \pm 0.0052_{\rm exp.}
\pm 0.0052_{\rm EW} \pm 0.0045_{\rm CKM}$
(independent of $d_3^{(0)}$ since it is NNLO method),
corresponding to
$\alpha_s(M^2_{\rm z}) = 0.1177 \pm 0.0007_{\rm exp.}
\pm 0.0009_{\rm EW+CKM}$.
However, this ECH method, applied directly to $r_{\tau}$,
appears to be unstable under 
the inclusion of the ${\rm N}^3 {\rm LO}$ information,
above all because the ECH renormalization scale
parameter for $r_{\tau}$ is dangerously low:
$\xi^2_{\rm ECH}\!\equiv\!\mu^2/m^2_{\tau} \approx 0.10$.
The ${\rm N}^3 {\rm LO}$ ECH approach to $r_{\tau}$,
with $d_3^{(0)}\!=\!25. \pm 10.$ (and using ${\rm N}^3 {\rm LO}$ TPS
$\beta$--functions of the ECH and ${\overline {\rm MS}}$ schemes)
thus gives very different results:
$\alpha_s(m^2_{\tau}) = 0.3373 \pm 0.0079_{\rm exp.}
\pm 0.0079_{\rm EW} \pm 0.0068_{\rm CKM} \pm 0.0192_{\delta d_3}$,
corresponding to
$\alpha_s(M^2_{\rm z}) = 0.1207 \pm 0.0009_{\rm exp.}
\pm 0.0012_{\rm EW+CKM} \pm 0.0020_{\delta d_3}$.

The authors of \cite{SE} used the diagonal $[2/2]$ Pad\'e
approximation to resum the ${\rm N}^3 {\rm LO}$ TPS
of $r_{\tau}$ (with $\xi^2\!=\!1$),
where the ${\rm N}^3 {\rm LO}$ coefficient $r_3$ of the series
was determined by the Asymptotic Pad\'e approximant method
(APAP) \cite{APAP}. They obtained 
$\alpha_s(m^2_{\tau}) = 0.314 \pm 0.010$.
The central value is significantly lower than our prediction
(\ref{res2}), although they used for the input the
values $r_{\tau} = 0.2048 \pm 0.0129$ where the
central value is considerably higher than that of our
input values (\ref{rtauexp}).

The results of the methods by both groups of authors \cite{KoKP,SE}
thus give in general lower predictions for
$\alpha_s$ than our method and the central value of the ALEPH method,
as also seen in Fig.~\ref{fig4}.
We wish to point out, however, that both groups of
authors of Refs.~\cite{KoKP,SE} 
applied resummation methods directly to the
observable $r_{\tau}$,
which is Minkowskian ($q^2\!=\!m^2_{\tau} > 0$).
We applied our resummation to the (Borel transform of the)
predominantly non--Minkowskian quantity $D(Q^2)$, i.e.,
we used the $y$--contour representation (\ref{contint}).\footnote{
The problematic Minkowskian region contribution
$q^2\!\equiv\!-Q^2 > 0$ ($y=\pm \pi$) in the contour
integral (\ref{contint}) is suppressed by the third power,
i.e., by the factor $(1\!+\!e^{{\rm i} y})^3$.}
Various authors \cite{KS,Mink1,Mink2} have suggested that
resummation techniques to (quasi)observables be used in the 
non--Minkowskian regions, because the physical
singularities appear on the Minkowskian axis 
($q^2\!\equiv\!-Q^2 > 0$).

The reason that the predictions of our method differ
significantly from those of the simpler CATPS $y$--contour
approach lies in the apparently important role of
the $b\!=\!2$ IR renormalon singularity of ${\widetilde D}(b)$
[see the ansatz in (\ref{BTd})--(\ref{rt3})] for the
quasianalytic continuation of ${\widetilde D}(b)$ and of
the Adler function $D(Q^2)$,
and consequently for the resummation of $r_{\tau}$
via the $y$--contour integration. 
This is so despite the fact that the $y$--contour
integration leads to a suppression of the
contributions from $b \approx 2$ (see also the last
paragraph of Sec.~\ref{sec:analysis}).

\section{Summary}
\label{sec:summary}

We presented a new method of determination of the
${\rm N}^3 {\rm LO}$ coefficient $d_3^{(0)}$ of the
Adler function $D(Q^2)$. The method makes use of the
known radiative correction to the $1/Q^4$--term
in the Operator Product Expansion (OPE) of $D(Q^2)$.
By requiring that the $Q$-dependence of the
ambiguity induced by the first nonzero infrared
renormalon of $D(Q^2)$ ($b\!=\!2$) is the same
as the $Q$--dependence of the OPE term
${\cal C}_4(a(Q^2))/Q^4$, the exact condition
(\ref{constraint2}) is obtained, which involves the
Borel transform of $D(Q^2)$. This condition,
in principle, would determine the coefficient
$d_3^{(0)}$. However, this condition has to
be evaluated at relatively large value $b\!=\!2$
of the Borel variable, and the present knowledge
of only two terms beyond the leading order leads to
significant uncertainties in the evaluation of this condition.
We solved this practical problem by applying
judicious conformal transformations $b\!=\!b(w)$ and Pad\'e
resummation techniques, thus improving
the convergence properties. The resulting value is
then: $d_3^{(0)} \approx 25. \pm 5.$ (at $N_f\!=\!3$),
but uncertainties of up to $\delta d_3^{(0)}= \pm 10.$
cannot be entirely excluded and were used in the
subsequent analyses of the $\tau$ inclusive hadronic decay ratio.

We wish to emphasize that our determination of $d_3^{(0)}$
is fundamentally different from the previous estimates in the
literature. The latter estimates were mainly based on reexpanding
the resummed (quasi--analytically continued) expressions for
$D(Q^2)$ in powers of the coupling parameter, 
thus relying on the assumption that a quasi--analytic continuation
of the NNLO truncated perturbation series of $D(Q^2)$
was efficient. However, this may only be true
if the main contribution to the coefficient $d_3^{(0)}$
comes from those higher order Feynman diagrams which
do not have new topological structures in comparison with
the lower order diagrams contributing to $d_2^{(0)}$
\cite{KS,Mink1,BLM}. In contrast, our relation (\ref{constraint2}),
and its evaluation, are not based just on the knowledge
of the NNLO truncated perturbation series, but also
on the knowledge of the first nonzero infrared
renormalon including its first radiative correction.
Therefore, it is possible that the resummations
of the expressions of (\ref{constraint2}) do not suffer
from the uncertainties about the
topologies of the Feynman diagrams.

We then used the obtained $d_3^{(0)}$, and the structure
of the Borel transform ${\widetilde D}(b)$ of $D(Q^2)$
near the first infrared renormalon at $b\!=\!2$,
and an optimal conformal transformation,
to evaluate the $\tau$ inclusive hadronic decay ratio 
$R_{\tau}$, or more specifically its massless QCD
reduced version $r_{\tau}$, 
via the contour integration method. Comparing the
obtained predictions with the precise experimental
data available now, we obtained the prediction
(\ref{res4}) for $\alpha_s(M^2_{\rm z})$, where the
estimated uncertainties from the method
(and RG evolution) do not surpass significantly
the uncertainties from the experimental data. 
All the uncertainties in (\ref{res4})--(\ref{res5})
are significantly lower than the uncertainties
in the present world average 
${\alpha}_s^{{\overline {\rm MS}}}(M_{\text{z}}^2) =
0.1173 \pm 0.0020$ by Ref.~\cite{Hinchliffe} and
$0.1184 \pm 0.0031$ by Ref.~\cite{Bethke}.
Furthermore, the central value (\ref{res4})
is by $0.0020$ and $0.0009$ higher than these two
world averages. 

In view of the present high precision
experimental data for the $R_{\tau}$ decay ratio,
we believe that the values of $\alpha_s(M^2_{\rm z})$
deduced from it should eventually serve as the
reference value for future tests of QCD via the
experimental measurements and theoretical analyses
of other QCD observables. For this, the theoretical
EW correction factor to $R_{\tau}$ should be investigated
further, and the present uncertainties in the
value of the CKM element $|V_{ud}|$ should be reduced.

\acknowledgments
G. C. is thankful to M.~Davier (ALEPH Collaboration)
for useful communication.
T.L. is thankful to Pyungwon Ko for useful discussions, and
was supported in part by the BK21 Core Program.
The work of G.C. was supported by the FONDECYT (Chile)
Grant No.~1010094.

\begin{appendix}

\section[]{Subtracting the quark mass effects}
\setcounter{equation}{0}

In order to be able to apply the massless QCD approach
to our analysis, we have to subtract the quark mass
($m_{u,d} \not= 0$) contributions from the reduced hadronic
$\tau$--decay width $r_{\tau}^{\rm V\!+\!A}(\triangle S\!=\!0)$.
The general expression for the latter quantity,
including the quark mass effects, is 
(see, for example, Refs.~\cite{BNP,rtau7})
\begin{eqnarray}
\lefteqn{
r_{\tau}^{{\rm V\!+\!A}}(\triangle S\!=\!0) \equiv 
\frac{ R_{\tau}^{{\rm V\!+\!A}}(\triangle S\!=\!0) }
{ 3 |V_{ud}|^2 (1 + {\delta}_{{\rm EW}}) } -
(1 + \delta_{\rm EW}^{\prime} )
}
\nonumber\\
&& = ( - \pi {\rm i} ) \int_{|s|=m^2_{\tau}}
\frac{ds}{s} \left( 1\!-\!\frac{s}{m^2_{\tau}} \right)^3
\left[ \left( 1 + \frac{s}{m^2_{\tau}} \right) D^{\rm L + T}(-s)
+ \frac{4}{3} D^{\rm L}(-s) \right] - 1 \ ,
\label{rtgen}
\end{eqnarray}
where the contour integration is counterclockwise in the complex $s$--plane,
and the general Adler functions $D^{\rm L + T}$ and $D^{\rm L}$
are expressed with the current--current correlation functions
\begin{eqnarray}
D^{\rm L + T}(-s) & = & - s \frac{d}{ds} \sum_{J=0,1}
\left( \Pi_{ud,V}^{(J)}(s) +\Pi_{ud,A}^{(J)}(s) \right) \ ,
\label{DLT} 
\\
D^{\rm L}(-s) & = & \frac{s}{m^2_{\tau}} \frac{d}{ds}
\left[ s \left( \Pi_{ud,V}^{(0)}(s) +\Pi_{ud,A}^{(0)}(s) \right) \right] \ ,
\label{DL}
\end{eqnarray}
where $\Pi^{\left(J\right)}_{ud,V/A}$
are components in the Lorentz decomposition 
\begin{equation}
\Pi^{\mu \nu}_{ud,V/A}(q) = ( - g^{\mu \nu} q^2 + q^\mu q^\nu )
\Pi^{(1)}_{ud,V/A}(q^2) + q^\mu q^\nu \Pi^{(0)}_{ud,V/A}(q^2) 
\label{Ldecomp}
\end{equation}
of the two--point
correlation functions $\Pi^{\mu \nu}_{ud,V/A}$
of the vector 
$V_{ud}^{\mu}\!=\!{\bar d} \gamma^{\mu} u$
and axial--vector
$A_{ud}^{\mu}\!=\!{\bar d} \gamma^{\mu} \gamma_5 u$
(color--singlet) currents
\begin{eqnarray}
- {\rm i} \Pi^{\mu \nu}_{ud,V}(q) &=& 
\int d^4 x \; e^{{\rm i} q\; \cdot\; x} 
\langle 0 | {\rm T} \lbrace V_{ud}^{\mu}(x) V_{ud}^{\nu}(0)^{\dagger} 
\rbrace | 0 \rangle \ ,
\label{Vcorr}
\\
- {\rm i} \Pi^{\mu \nu}_{ud,A}(q) &=& 
\int d^4 x \; e^{{\rm i} q \; \cdot \; x} 
\langle 0 | {\rm T} \lbrace A_{ud}^{\mu}(x) A_{ud}^{\nu}(0)^{\dagger} 
\rbrace | 0 \rangle \ .
\label{Acorr}
\end{eqnarray}
In the massless quark limit ($m_{u,d} \to 0$), $D^{\rm L}(s)$ vanishes,
the perturbative vector and axial--vector contributions in
$D^{\rm L + T}$ become equal
and $D^{\rm L + T}(-s) \to (1 + D(-s))/(2 \pi^2)$,
where $D(Q^2)$ is the canonically normalized massless
Adler function (\ref{ddef}) with the perturbative
expansion (\ref{expansion}).\footnote{
Usually in the literature (e.g., see Refs.~\cite{rtau7}),
the ($ud$) Adler functions $D^{\rm L+T}$ and $D^{\rm L}$
(\ref{DLT})--(\ref{DL}) include by convention the
additional CKM factor $|V_{ud}|^2$.}
In order to apply the massless QCD analysis to
the measured observable (\ref{rtgen}), we have to
subtract from it the quark mass ($m_{u,d} \not= 0$)
contributions. These are largely the $m_{\pi} \not= 0$
contributions from the pion ($\pi^-$) pole. 
The pion pole contributes to the axial--current 
correlation functions $\Pi^{(J)}_{ud,A}(s)$.
Using PCAC, these contributions can be obtained, and they lead to
the corresponding contributions in the Adler functions: 
$D^{\rm L}(-s; \pi) = 2 f^2_{\pi} m^2_{\pi} s/(s - m^2_{\pi})^2/m^2_{\tau}$ 
and
$D^{\rm L + T}(-s; \pi) = - 2 f^2_{\pi} s/(s - m^2_{\pi})^2$.
This leads via (\ref{rtgen}) to the following estimate of the
pion pole contribution to $r_\tau^{\rm V+A}(\Delta S=0)$:
\begin{equation}
r_{\tau}^{{\rm V\!+\!A}}(\triangle S\!=\!0; \pi) =
\frac{8 \pi^2 f^2_{\pi}}{m^2_{\tau}}
\left( 1 - \frac{m^2_{\pi}}{m^2_{\tau}} \right)^2
\approx 0.2135*(1 - 0.0123) \approx 0.2109 \ .
\label{rtpi}
\end{equation}
Here, we employed the known values \cite{PDG2000}:
$f_{\pi} = 92.4 \pm 0.3$ MeV, $m_{\pi^-} = 139.6$ MeV,
$m_{\tau} = 1777$ MeV.
In order to check whether the framework leading to (\ref{rtpi})
is realistic, we may calculate from here the branching ratio
for $\tau^- \to \pi^- \nu_{\tau}$
\begin{eqnarray}
B(\tau^- \to \pi^- \nu_{\tau}) &=& 
\frac{ \Gamma(\tau^- \to \pi^- \nu_{\tau}) }
{ \Gamma(\tau^- \to e^- {\overline \nu}_e \nu_{\tau} )}
B(\tau^- \to e^- {\overline \nu}_e \nu_{\tau} )
\nonumber\\
& = & R_{\tau}(\pi) B_e \approx 3 |V_{ud}|^2 
r_{\tau}^{{\rm V\!+\!A}}(\triangle S\!=\!0; \pi) 
\approx 0.1072 \ ,
\label{Bpi}
\end{eqnarray}
where we used for the branching ratio
$B_e \equiv B(\tau^- \to e^- {\overline \nu}_e \nu_{\tau} )$
the middle value of the world average \cite{PDG2000} 
$B_e = 0.1783$, and $|V_{ud}| = 0.9749$.
On the other hand, the measured branching ratio
for $\tau^- \to \pi^- \nu_{\tau}$ is
$B(\tau^- \to \pi^- \nu_{\tau}) = 0.1109 \pm 0.0012$
\cite{PDG2000}. The value (\ref{Bpi}), obtained from
the PCAC--motivated approach (\ref{rtpi}), thus differs 
less than $4 \%$ from the actual prediction.

We can now read from the expression (\ref{rtpi}) the
quark mass ($m_{u,d} \not= 0$, i.e., $m_{\pi} \not= 0$) 
contribution to $r_{\tau}^{{\rm V\!+\!A}}(\triangle S\!=\!0)$
\begin{equation}
\delta r_{\tau}^{{\rm V\!+\!A}}(\triangle S\!=\!0)_{m_{\pi} \not= 0} =
- \frac{16 \pi^2 f^2_{\pi} m^2_{\pi}}{m^4_{\tau}}
\left( 1 - \frac{m^2_{\pi}}{2 m^2_{\tau}} \right)
\approx - 0.0026 \ .
\label{drtmpi}
\end{equation}
However, we can go somewhat beyond the approximation made so far
in calculating this contribution. In the Operator Product Expansion
(OPE) approach to $R_{\tau}$ ratio, as given in \cite{BNP},
the largest quark mass contributions are of dimension
$d=4$ ($\propto\!1/m^4_{\tau}$, quark condensate contributions)
\begin{eqnarray}
\lefteqn{
\delta r_{\tau}^{{\rm V\!+\!A}}(\triangle S\!=\!0)_{m_{u,d}\not=0}
\approx 16 \pi^2 
\frac{ (m_u\!+\!m_d)\langle {\bar q} q \rangle }{m^4_{\tau}}
\left[ 1 + \frac{23}{8} 
\left( \frac{ \alpha_s(m^2_{\tau}) }{\pi} \right)^2 \right]
}
\label{drtmud1}
\\
&\approx & - \frac{16 \pi^2 f^2_{\pi} m^2_{\pi}}{m^4_{\tau}}
\left[ 1 + \frac{23}{8} 
\left( \frac{ \alpha_s(m^2_{\tau}) }{\pi} \right)^2 \right]
\approx - 0.0027 \ .
\label{drtmud2}
\end{eqnarray}
In (\ref{drtmud1}) we denoted 
$\langle {\bar q} q \rangle \equiv \langle {\bar u} u \rangle
\approx \langle {\bar d} d \rangle$. The renormalization scale
in this quantity and in $m_u$ and $m_d$ in 
(\ref{drtmud1})--(\ref{drtmud2}) can be taken to be
$\mu \approx m_{\tau}$. In (\ref{drtmud2}) we took into
account the PCAC relation $(m_u\!+\!m_d)\langle {\bar q} q \rangle
\approx - f^2_{\pi} m^2_{\pi}$. There are corrections to the
expression (\ref{drtmud2}) of the order 
$\sim\!m^2_{u,d}/m^2_{\tau}$, i.e., of the order of the OPE
$d\!=\!2$ terms which can reach, at most, values $\sim\!10^{-4}$.
Comparing the previous 
pion pole expression (\ref{drtmpi}) with the OPE expression
(\ref{drtmud2}), we see that the latter apparently represents
a slight improvement since it includes the radiative corrections.
In obtaining the number
(\ref{drtmud2}), we further used the value
$\alpha_s(m^2_{\tau},{\overline {\rm MS}}) \approx 0.32$.

The OPE approach of \cite{BNP} includes other
nonperturbative terms contributing to $r_{\tau}$, which 
do not stem from quark masses:
the $d=4$ gluon condensate, and the $d=6$ term.
The latter term could be large, but it has also comparably
large uncertainties \cite{BNP}. The gluon condensate
contribution to $r_{\tau}$ in the OPE approach is $\alpha_s$--suppressed.
The ALEPH analysis \cite{ALEPH1}
indicates that these $d=4,6$ nonperturbative contributions
are consistent with the value zero.

When subtracting the quark mass contributions
(\ref{drtmud1})--(\ref{drtmud2}) from (\ref{rtgen}), we
end up with the massless QCD observable
\begin{eqnarray}
r_{\tau} &\equiv& 
r_{\tau}^{{\rm V\!+\!A}}(\triangle S\!=\!0; m_{u,d} =0) =
r_{\tau}^{{\rm V\!+\!A}}(\triangle S\!=\!0) -
\delta r_{\tau}^{{\rm V\!+\!A}}(\triangle S\!=\!0)_{m_{u,d}\not=0}
\label{rtmud1=0}
\\
& = & - \frac{ {\rm i}}{2 \pi} \int_{|s|=m^2_{\tau}}
\frac{ds}{s} \left( 1\!-\!\frac{s}{m^2_{\tau}} \right)^3
\left( 1 + \frac{s}{m^2_{\tau}} \right) D(-s) \ ,
\label{rtmud2=0}
\end{eqnarray}
where the integration is counterclockwise, and the
canonically normalized massless Adler function $D(Q^2\!\equiv\!-s)$
(\ref{ddef}), (\ref{expansion}), was introduced according to
the aforementioned limiting procedure: 
$D^{\rm L}(s) \to 0$ and $D^{\rm L + T}(-s) \to (1 + D(-s))/(2 \pi^2)$
(when $m_{u,d} \to 0$).

\end{appendix}

\begin{thebibliography}{99}



\bibitem{parisi} 
G. Parisi, Phys. Lett. B {\bf 76}, 65 (1978);
Nucl. Phys. B {\bf 150}, 163 (1979).

\bibitem{hooft}
G. `t  Hooft, The whys of subnuclear physics (Erice, 1977), 
ed. A. ~Zichichi (Plenum, New York, 1977).

\bibitem{david}
F. David, Nucl. Phys. B {\bf 234}, 237 (1984);
Nucl. Phys. B {\bf 263}, 637 (1986).

\bibitem{KS}
A.~L.~Kataev and V.~V.~Starshenko, 
Mod. Phys. Lett. A {\bf 10}, 235 (1995).

\bibitem{BBB}
M.~Beneke and V. M. Braun, Phys. Lett. B {\bf 348}, 513 (1995);
P. Ball, M. Beneke, and V. M. Braun, Nucl. Phys. {\bf B452}, 563 (1995).


\bibitem{Neubert}
M.~Neubert, Nucl. Phys. {\bf B463}, 511 (1996).

\bibitem{LTM}
C.~N.~Lovett-Turner and C. J. Maxwell, 
Nucl. Phys. {\bf B452}, 188 (1995).

\bibitem{tlee}
T. Lee, Phys. Rev. D {\bf 56}, 1091 (1997); 
Phys. Lett. B {\bf 462},1 (1999).

\bibitem{mueller} 
A. H. Mueller, Nucl. Phys. {\bf B250}, 327 (1985).

\bibitem{beneke} 
M. Beneke, Phys. Lett. B {\bf 307}, 154 (1993).

\bibitem{soper}
D. Soper and L.R. Surguladze,
Phys. Rev. D {\bf 54}, 4566 (1996).

\bibitem{wilson} 
K.G. Chetyrkin, S.G. Gorishny and 
V.P. Spiridonov, Phys. Lett. B {\bf 160}, 149 (1985).

\bibitem{BNP}
E.~Braaten, S.~Narison, and A.~Pich, Nucl. Phys. {\bf B373}, 581 (1992).

\bibitem{mueller1} 
A. H. Mueller, Phys. Lett. B {\bf 308}, 355 (1993).
  
\bibitem{caprini} 
I. Caprini and J. Fischer, Phys. Rev. D {\bf 60}, 054014 (1999). 

\bibitem{coeffs1}
K.~G.~Chetyrkin, A.~L.~Kataev, and F.~V. Tkachev,
Phys. Lett. B {\bf 85}, 277 (1979);
M. Dine and J. Sapirstein, Phys. Rev. Lett. {\bf 43}, 668 (1979);
W. Celmaster and R. Gonsalves, Phys. Rev. Lett. {\bf 44}, 560 (1980).

\bibitem{coeffs2}
S.~G.~Gorishny, A.~L.~Kataev, and  S.~A.~Larin,
Phys. Lett. B {\bf 259}, 144 (1991);
L.~R. Surguladze and M.~A. Samuel, Phys. Rev. Lett. {\bf 66}, 560 (1991).

\bibitem{JS}
U.~D.~Jentschura and G. Soff, J. Phys. A {\bf 34}, 1451 (2001).

\bibitem{Pade}
George A. Baker, Jr. and Peter Graves--Morris,
{\it Pad\'e Approximants\/}, 2nd edition, 746 pp.
(Encyclopedia of Mathematics and Its Applications, Vol.~59),
edited by Gian-Carlo Rota
(Cambridge University Press, 1996).

\bibitem{ECH}
G.~Grunberg, Phys. Lett. B {\bf 95}, 70 (1980),
{\em ibid\/} B {\bf 110}, 501(E) (1982);
{\em ibid\/} B {\bf 114}, 271 (1982);
Phys. Rev. D {\bf 29}, 2315 (1984).

\bibitem{KKP}
N.~V.~Krasnikov, Nucl. Phys. {\bf B192}, 497 (1981);
A.~L. Kataev, N.~V. Krasnikov, and A.~A. Pivovarov,
Nucl. Phys. {\bf B198}, 508 (1982).

\bibitem{Gupta}
A.~Dhar and V.~Gupta, Phys. Rev. D {\bf 29}, 2822 (1984);
V.~Gupta, D. V. Shirkov, and O. V. Tarasov, 
Int. J. Mod. Phys. A {\bf 6}, 3381 (1991).

\bibitem{PMS}
P.~M.~Stevenson, Phys. Rev. D {\bf 23}, 2916 (1981);


\bibitem{PMS2}
P.~M.~Stevenson, Phys. Lett. B {\bf 100}, 61 (1981);
Nucl. Phys. {\bf B203}, 472 (1982).

\bibitem{rtau0}
C.~S.~Lam and T.~M. Yan, Phys. Rev. D {\bf 16}, 703 (1977).

\bibitem{rtau1}
K.~Schilcher and M.~D.~Tran, Phys. Rev. D {\bf 29}, 570 (1984).

\bibitem{rtau2}
S.~Narison and A.~Pich, Phys. Lett. B {\bf 211}, 183 (1988).

\bibitem{rtau3}
E.~Braaten, Phys. Rev. Lett. {\bf 60}, 1606 (1988);
Phys. Rev. D {\bf 39}, 1458 (1989).

\bibitem{rtau4}
A.~Pich, Hadronic tau-decays and QCD, in Proc. Workshop on
Tau Lepton Physics, Orsay, France, September 1990, 
eds. M.~Davier and B. Jean-Marie
(Ed. Fronti\`eres, Paris, 1991), p.321 
(see also preprint CERN-TH-5940-90, KEK Library).
 
\bibitem{rtau6}
F.~Le Diberder and A.~Pich, Phys. Lett. B {\bf 286}, 147 (1992).

\bibitem{rtau7}
A.~Pich and J.~Prades, JHEP {\bf 06}, 013 (1998); 
{\em ibid.\/} {\bf 10}, 004 (1999).

\bibitem{MS}
W.~Marciano and A.~Sirlin, Phys. Rev. Lett. {\bf 56}, 22 (1986);
{\em ibid.\/} {\bf 61}, 1815 (1988).

\bibitem{BL}
E.~Braaten and C.~S.~Li, Phys. Rev. D {\bf 42}, 3888 (1990).

\bibitem{PDG2000}
Particle Data Book 2000, Eur. Phys. J. C {\bf 15}, 695 (2000).

\bibitem{ALEPH1}
ALEPH Collaboration, Eur. Phys. J. C {\bf 4}, 409 (1998).

\bibitem{ALEPH2}
ALEPH Collaboration, Eur. Phys. J. C {\bf 11}, 599 (1999); 
{\em ibid.\/} {\bf 10}, 1 (1999).

\bibitem{Davier}
M.~Davier, hep-ex/9912044.

\bibitem{OPAL}
OPAL Collaboration, Eur. Phys. J. C {\bf 13}, 197 (2000).

\bibitem{CLEO}
CLEO Collaboration, Phys. Rev. D {\bf 60}, 112002 (1999).

\bibitem{Davier2}
M.~Davier, private communication; and talk given at the
Sixth International Workshop on Tau Lepton Physics
(Tau'2000 Workshop), September 2000, Victoria, Canada
(to appear on http://tau2000.phys.uvic.ca/).

\bibitem{RVL}
T.~van Ritbergen, J.~A.~M.~Vermaseren and S.~A.~Larin,
Phys. Lett. B {\bf 400}, 379 (1997).

\bibitem{CK1}
G.~Cveti\v c, Phys. Lett. B {\bf 486}, 100 (2000);
G.~Cveti\v c and R.~K\"ogerler,
Phys. Rev. {\bf D63} (2001) 056013.

\bibitem{CK2}
G.~Cveti\v c, Nucl. Phys. {\bf B517}, 506 (1998);
Phys. Rev. D {\bf 57}, R3209 (1998).
G.~Cveti\v c and R.~K\"ogerler, Nucl. Phys. {\bf B522}, 396 (1998).





\bibitem{GKP}
S.~Groote, J.~G.~K\"orner, and A. A. Pivovarov,
Phys. Lett. B {\bf 407}, 66 (1997).

\bibitem{CKT}
K.~G.~Chetyrkin, A.~L.~Kataev, and  F. V. Tkachev,
Nucl. Phys. {\bf B174}, 345 (1980).

\bibitem{Raczka}
P.~A.~R\c aczka, Phys. Rev. D {\bf 57}, 6862 (1998).

\bibitem{BBK}
M.~Beneke, V.~M.~Braun, and N. Kivel,
Phys. Lett. B {\bf 404}, 315 (1997).

\bibitem{CKS}
K.~G. Chetyrkin, B.~A. Kniehl, and M. Steinhauser,
Phys. Rev. Lett. {\bf 79}, 2184 (1997).

\bibitem{Hinchliffe}
I.~Hinchliffe and A.~V.~Manohar, hep-ph/0004186.

\bibitem{Bethke}
S. Bethke, J. Phys. G {\bf 26}, R27 (2000).

\bibitem{duality}
B.~Chibisov, R.~D.~Dikeman, M. Shifman, and N.~G. Uraltsev,
Int. J. Mod. Phys. A {\bf 12}, 2075 (1997);
I. Bigi, M. Shifman, N. Uraltsev, and A. Vainstein,
Phys. Rev. D {\bf 59}, 054011 (1999).

\bibitem{CNZ}
K.~G.~Chetyrkin, S. Narison, and V.~I. Zakharov,
Nucl. Phys. {\bf B550}, 353 (1999);

\bibitem{HRS}
S.~J. Huber, M. Reuter, and M.~G. Schmidt,
Phys. Lett. B {\bf 462}, 158 (1999).

\bibitem{DoSch}
C.~A.~Dominguez and K. Schilcher,
Phys. Rev. D {\bf 61}, 114020 (2000).

\bibitem{gr1}
K.~Maltman, Phys. Lett. B {\bf 440}, 367 (1998).

\bibitem{gr2}
C.~A.~Dominguez and K. Schilcher,
Phys. Lett. B {\bf 448}, 93 (1999).

\bibitem{LDP2}
F.~Le Diberder and A. Pich, Phys. Lett. B {\bf 289}, 165 (1992).

\bibitem{GKPMPLA}
S.~Groote, J.~G. K\"orner, and A.~A. Pivovarov,
Mod. Phys. Lett. A {\bf 13}, 637 (1998).

\bibitem{KoKP}
J.~G.~K\"orner, F. Krajewski, and A. A. Pivovarov,
Phys. Rev. D {\bf 63}, 036001 (2001).

\bibitem{SE}
T.~G.~Steele and V. Elias, Mod. Phys. Lett. A {\bf 13}, 3151 (1998).

\bibitem{APAP}
J.~Ellis, M. Karliner, and  M.~A. Samuel,
Phys. Lett. B {\bf 400}, 176 (1997);
J. Ellis, I. Jack, D.~R.~T. Jones, M. Karliner, and M.~A. Samuel,
Phys. Rev. D {\bf 57}, 2665 (1998).

\bibitem{Mink1}
A.~L.~Kataev and V. V. Starshenko,
Phys. Rev. D {\bf 52}, 402 (1995);

\bibitem{Mink2}
K. G. Chetyrkin, B. A. Kniehl, and A. Sirlin,
Phys. Lett. B {\bf 402}, 359 (1997).

\bibitem{BLM}
S.~J.~Brodsky, G.~P. Lepage, and P.~B. Mackenzie,
Phys. Rev. D {\bf 28}, 228 (1983).

\end{thebibliography}
\end{document}